\documentstyle[aps, prb, epsf, floats]{revtex}

\begin{document}
\author{H. N\'{e}lisse, C. Bourbonnais$^{\dagger }$, H. Touchette, Y.M. Vilk and
A.-M. S. Tremblay$^{\dagger }$}
\title{Spin susceptibility of interacting electrons in one dimension: Luttinger
liquid and lattice effects.}
\address{D\'{e}partment de physique and Centre de recherche en physique du\\
solide\\
$^{\dagger }$Institut canadien de recherches avanc\'{e}es\\
Universit\'{e} de Sherbrooke, Sherbrooke, Qu\'{e}bec, Canada J1K 2R1}
\date{\today }
\maketitle

\begin{abstract}
The temperature-dependent uniform magnetic susceptibility of interacting
electrons in one dimension is calculated using several methods. At low
temperature, the renormalization group reaveals that the Luttinger liquid
spin susceptibility $\chi \left( T\right) $ approaches zero temperature with
an infinite slope in striking contrast with the Fermi liquid result and with
the behavior of the compressibility in the absence of umklapp scattering.
This effect comes from the leading marginally irrelevant operator, in
analogy with the Heisenberg spin $1/2$ antiferromagnetic chain. Comparisons
with Monte Carlo simulations at higher temperature reveal that
non-logarithmic terms are important in that regime. These contributions are
evaluated from an effective interaction that includes the same set of
diagrams as those that give the leading logarithmic terms in the
renormalization group approach. Comments on the third law of thermodynamics
as well as reasons for the failure of approaches that work in higher
dimensions are given.
\end{abstract}


\section{Introduction}

There are a number of organic conductors, including the Bechgaard salts for
example, for which the distinctive behavior of one-dimensional interacting
electrons is observed over a wide range of temperature.\cite{Revue,Wzietek93}
Among the characteristics observed is the asymptotic low-frequency,
long-wavelength electronic behavior which, in one dimension, belongs to the
universality class of Luttinger liquids. This universality class plays in
one-dimension the role of Landau Fermi liquid theory in higher dimension,
providing a framework to understand the occurence of power law behavior,
spin-charge separation and various other characteristics of one-dimensional
systems \cite{Voit95}.

While much is known about the predictions of Luttinger liquid theory,
non-singular quantities, such as the uniform magnetic spin susceptibility $%
\chi ,$ are not completely understood theoretically. Experimentally, $\chi
\left( T\right) $ can be accurately measured as a function of temperature
using a number of techniques, including Knight shift in Nuclear Magnetic
Resonance (NMR) experiments. In higher dimension, the theoretical situation
for $\chi \left( T\right) $ is clear within Landau Fermi liquid theory. The
prediction for this key quantity is a Pauli-like susceptibility whose
absolute value is enhanced by interactions. For Luttinger liquid theory in
one dimension pioneering work was done by Dzyaloshinskii and Larkin.\cite{DL}
and by Lee {\it et al.}\cite{Lee} A few years ago, Bourbonnais\cite
{BourbonParis}, using the renormalization group, readdressed this problem.
The conclusion that emerges from these works is that the temperature
dependence of $\chi \left( T\right) $ remains important up to $T=0,$
contrary to the Fermi liquid result in high dimension where temperature
dependence shows up only on the scale of the Fermi energy. The
renormalization group approach is expected to give the correct asymptotic
low-temperature behavior of the Luttinger liquid. To obtain quantitative
confirmation of these results from a specific microscopic model, it is
customary to study the Hubbard model that has an exact Bethe {\it ansatz}
solution in one dimension and belongs to the Luttinger-liquid universality
class. Despite the exact solution, the Bethe {\it ansatz }magnetic
susceptibility must be computed numerically.\cite{BetheChi} Additional
results have been obtained recently by J\"{u}ttner et al.\cite{Juttner}
through a quantum transfer matrix method, by Mila and Penc through world
line quantum Monte Carlo simulations,\cite{Mila} and by Moukouri\cite
{Moukouri} through Density Matrix Renormalization Group techniques. It is
important to notice that the low temperature limit is difficult to reach
numerically using any of the available numerical approaches, including Bethe 
{\it ansatz}.

In this paper, following Refs.\cite{BourbonParis} and\cite{Nelisse}, we
derive the renormalization group (RG) prediction for the
temperature-dependent uniform magnetic susceptibility of the {\em g-}ology
Hamiltonian. We show that the uniform magnetic susceptibility $\chi \left(
T\right) $ of this Luttinger liquid approaches zero temperature with an
infinite slope, in analogy with the spin $1/2$ antiferromagnetic Heisenberg
chain\cite{Affleck94,Dumoulin96}. This effect comes from backscattering of
right- and left-moving electrons, which is the leading marginally irrelevant
operator. We show that this phenomenon appears in a temperature range where
no numerical calculation has been done yet. We also obtain predictions at
higher temperature by using numerical simulations done on the Hubbard
Hamiltonian at quarter filling. To obtain {\it quantitative} agreement with
the simulations, we find it necessary to include non-logarithmic
contributions that are beyond the RG treatment. This is done by using an
approach inspired from the Kanamori\cite{Kanamori}-Brueckner\cite{Brueckner}
theory valid in higher dimension\cite{ChenLiBourbonnais}. The subset of
diagrams that must be resummed in one dimension is suggested by the RG and
it does differ from the subset used in higher dimension. The reasons for the
failure of higher-dimensional approaches are also discussed.

The {\em g-}ology and Hubbard Hamiltonians are introduced in Sec.II. In Sec.
III, we present the results of numerical calculations and in Sec. IV the RG
calculation, including the prediction of the infinite slope as $T\rightarrow
0$. Sec. V discusses the comparison between numerical results, RG prediction
and the reasons for the failure of higher-dimensional approaches. We conlude
with a summary of our main results and general comments on the range of
applicability of Luttinger-liquid theory.

\section{Hubbard model and connection with {\em g-}ology}

The simulations are done for the Hubbard model 
\begin{equation}
H=-t\sum_{<ij>\sigma }\left( c_{i\sigma }^{\dagger }c_{j\sigma }+c_{j\sigma
}^{\dagger }c_{i\sigma }\right) +U\sum_{i}n_{i\uparrow }n_{i\downarrow
\,\,\,\,}  \label{Hubbard}
\end{equation}
where units of energy are chosen such that the hopping matrix element $t$
equals unity in the simulations. The creation (annihilation) operators $%
c_{i\sigma }^{\dagger }\left( c_{i\sigma }\right) $ create (annihilate)
electrons of spin $\sigma $ in the orbital located on site $i$ with position 
$r_{i}.$ Only nearest-neighbor hopping is allowed. The last term, with the
usual occupation number operators $n_{i\uparrow }=$ $c_{i\sigma }^{\dagger
}c_{i\sigma }$, represents the short-range repulsion $U,$ felt by the
electrons when they occupy the same orbital at site $i$.

As is well known,\cite{Classiques} there are logarithmic divergences in the
perturbative treatment of the Hubbard model in one dimension. These can be
handled directly by infinite resummations of parquet diagrams,\cite
{Classiques} or most easily by a renormalization group treatment.\cite
{Solyom79},\cite{Emery},\cite{Firsov},\cite{BourbonnaisCaron} In this case,
the Hubbard Hamiltonian is not a fixed-point Hamiltonian. It is necessary to
consider the renormalization group flows in a more general space of
Hamiltonians called {\em g-}ology Hamiltonian. In the rest of this section,
we recall how to cast the Hubbard Hamiltonian as a special case of {\em g-}%
ology. First, it is useful to rewrite it in the form 
\begin{equation}
H=-t\sum_{<ij>\sigma }\left( c_{i\sigma }^{\dagger }c_{j\sigma }+c_{j\sigma
}^{\dagger }c_{i\sigma }\right) +\frac{U}{2}\sum_{i\sigma \sigma ^{\prime
}}c_{i\sigma }^{\dagger }c_{i\sigma ^{\prime }}^{\dagger }c_{i\sigma
^{\prime }}c_{i\sigma }.
\end{equation}
which, compared with Eq.(\ref{Hubbard}) contains an additional term that can
be absorbed in a chemical potential shift. The {\em g-}ology Hamiltonian
being defined in Fourier space, we take 
\begin{equation}
c_{\sigma }\left( k\right) =\frac{1}{\sqrt{L}}\sum_{j=1}^{L}e^{-ikr_{j}}c_{j%
\sigma }\quad ;\quad c_{\sigma }^{\dagger }\left( k\right) =\frac{1}{\sqrt{L}%
}\sum_{j=1}^{L}e^{ikr_{j}}c_{j\sigma }^{\dagger }
\end{equation}
where we have chosen unity for the lattice spacing so that the number of
sites and the system size are both equal to $L$. Using these variables, and
neglecting {\em umklapp} processes, we can write, 
\begin{equation}
H=\sum_{k,\sigma }\left( -2t\cos k\right) c_{\sigma }^{\dagger }\left(
k\right) c_{\sigma }\left( k\right) +\frac{U}{2L}\sum_{k,k^{\prime
},q,\sigma ,\sigma ^{\prime }}c_{\sigma }^{\dagger }\left( k\right)
c_{\sigma ^{\prime }}^{\dagger }\left( k^{\prime }\right) c_{\sigma ^{\prime
}}\left( k^{\prime }+q\right) c_{\sigma }\left( k-q\right) .
\label{HubbardFourier}
\end{equation}

In recent versions of the renormalization group\cite{Dumoulin96}, the full
cosine dispersion relation can be taken into account, but in the more usual
version that we consider here, the dispersion relation is linearized around
the two Fermi points $\pm k_{F}$, and one considers only scatterings around
and between these points. To rewrite the Hubbard Hamiltonian Eq.(\ref
{HubbardFourier}) in a way that highlights the processes that are allowed by
the Pauli principle, the sum over momentum transfers $q$ is divided also in
three pieces: $q\approx 0$, and $q\approx \pm 2k_{F}$. Furthermore, one
introduces a lower index $p$ to the creation-annihilation operators that,
for the moment, just indicates if the allowed particle momenta are mostly
around the $+k_{F}$ Fermi point (right-moving $\left( +\right) $), or around
the $-k_{F}$ Fermi point or (left-moving $\left( -\right) $). This
rearrangement gives, after one allows the $k$ sums to run from $%
-k_{0}+pk_{F} $ to $k_{0}+pk_{F}$ with $k_{0}$ a cut-off wave vector of the
order of $k_{F} $. 
\begin{eqnarray}
H &\approx &\sum_{k,\sigma ,p}\epsilon _{p}\left( k\right) c_{p,\sigma
}^{\dagger }\left( k\right) c_{p,\sigma }\left( k\right)  \nonumber \\
&&+\frac{U}{2L}\sum_{\sigma ,\sigma ^{\prime }=\uparrow }^{\downarrow
}\sum_{\left( p,p^{\prime }=\pm \right) }\sum_{\left( k=-k_{0}+pk_{F}\right)
}^{k_{0}+pk_{F}}\sum_{\left( k^{\prime }=-k_{0}+p^{\prime }k_{F}\right)
}^{k_{0}+p^{\prime }k_{F}}\sum_{q}c_{p,\sigma }^{\dagger }\left( k\right)
c_{p^{\prime },\sigma ^{\prime }}^{\dagger }\left( k^{\prime }\right)
c_{p^{\prime },\sigma ^{\prime }}\left( k^{\prime }+q\right) c_{p,\sigma
}\left( k-q\right)  \nonumber \\
&&+\frac{U}{2L}\sum_{\sigma ,\sigma ^{\prime }=\uparrow }^{\downarrow
}\sum_{\left( k=-k_{0}+k_{F}\right) }^{k_{0}+k_{F}}\sum_{\left( k^{\prime
}=-k_{0}-k_{F}\right) }^{k_{0}-k_{F}}\sum_{q}c_{+,\sigma }^{\dagger }\left(
k\right) c_{-,\sigma ^{\prime }}^{\dagger }\left( k^{\prime }\right)
c_{+,\sigma ^{\prime }}\left( k^{\prime }+q+2k_{F}\right) c_{-,\sigma
}\left( k-q-2k_{F}\right)  \nonumber \\
&&+\frac{U}{2L}\sum_{\sigma ,\sigma ^{\prime }=\uparrow }^{\downarrow
}\sum_{\left( k^{\prime }=-k_{0}+k_{F}\right) }^{k_{0}+k_{F}}\sum_{\left(
k=-k_{0}-k_{F}\right) }^{k_{0}-k_{F}}\sum_{q}c_{-,\sigma }^{\dagger }\left(
k\right) c_{+,\sigma ^{\prime }}^{\dagger }\left( k^{\prime }\right)
c_{-,\sigma ^{\prime }}\left( k^{\prime }+q-2k_{F}\right) c_{+,\sigma
}\left( k-q+2k_{F}\right)  \label{Intermediate}
\end{eqnarray}
where the linearized dispersion relation is 
\begin{equation}
\epsilon _{p}\left( k\right) =pv_{F}\left( k-pk_{F}\right) \quad ;\quad
v_{F}\equiv 2t\sin k_{F}.  \label{dispersion}
\end{equation}
The restrictions on momentum transfer $q$ should normally be set to avoid
double counting various scattering processes. However, it is simpler to
introduce additional states that linearly extrapolate the right and
left-moving electron dispersion relations. In otherwords, strikctly speaking
we should have $c_{+,\sigma }\left( k\right) =c_{\sigma }\left( k\right) $
for $0\leq k<\pi $ and $c_{-,\sigma }\left( k\right) =c_{\sigma }\left(
k\right) \ $for $-\pi \leq k<0$ while, instead, we add states in such a way $%
k$ runs from $-\infty $ to $+\infty $ for both cases $p=\pm 1$. Each of
these sets, $p=\pm 1$, are defined as a ``branch'' with the corresponding
dispersion relation $\epsilon _{p}\left( k\right) =pv_{F}\left(
k-pk_{F}\right) .$ The added unphysical states should not contribute
appreciably because of the large energy denominators,the Pauli principle and
the cut-offs in the sums over $k$ and $k^{\prime }$ that regularize
perturbation theory. Hence, in the weak to intermediate coupling regime only
scatterings near the Fermi surface are important so we can assume that the
sum over $q$ that is left for each of the three pieces is free to run from $%
-\infty $ to $+\infty $. In the {\em g-}ology notation, the last two terms
of Eq.(\ref{Intermediate}) are regrouped into $2k_{F}$ scatterings with an
interaction constant $g_{1}$, while the first interaction term is divided
into $q\approx 0$ scatterings on the same branch $\left( g_{4}\right) $, and 
$q\approx 0$ scatterings between two different branches $\left( g_{2}\right) 
$, namely 
\begin{eqnarray}
H_{g} &=&\sum_{k,\sigma ,p}\epsilon _{p}\left( k\right) c_{p,\sigma
}^{\dagger }\left( k\right) c_{p,\sigma }\left( k\right)  \nonumber \\
&&+\frac{g_{1}}{2L}\sum_{q}\sum_{\sigma ,\sigma ^{\prime }=\uparrow
}^{\downarrow }\sum_{p=\pm }\sum_{\left( k=-k_{0}+pk_{F}\right)
}^{k_{0}+pk_{F}}\sum_{\left( k^{\prime }=-k_{0}-pk_{F}\right)
}^{k_{0}-pk_{F}}c_{p,\sigma }^{\dagger }\left( k\right) c_{-p,\sigma
^{\prime }}^{\dagger }\left( k^{\prime }\right) c_{p,\sigma ^{\prime
}}\left( k^{\prime }+q+2pk_{F}\right) c_{-p,\sigma }\left( k-q-2pk_{F}\right)
\nonumber \\
&&+\frac{g_{2}}{2L}\sum_{q}\sum_{\sigma ,\sigma ^{\prime }=\uparrow
}^{\downarrow }\sum_{p=\pm }\sum_{\left( k=-k_{0}+pk_{F}\right)
}^{k_{0}+pk_{F}}\sum_{\left( k^{\prime }=-k_{0}-pk_{F}\right)
}^{k_{0}-pk_{F}}c_{p,\sigma }^{\dagger }\left( k\right) c_{-p,\sigma
^{\prime }}^{\dagger }\left( k^{\prime }\right) c_{-p,\sigma ^{\prime
}}\left( k^{\prime }+q\right) c_{p,\sigma }\left( k-q\right)  \nonumber \\
&&+\frac{g_{4}}{2L}\sum_{q}\sum_{\sigma ,\sigma ^{\prime }=\uparrow
}^{\downarrow }\sum_{p=\pm }\sum_{\left( k=-k_{0}+pk_{F}\right)
}^{k_{0}+pk_{F}}\sum_{\left( k^{\prime }=-k_{0}+pk_{F}\right)
}^{k_{0}+pk_{F}}c_{p,\sigma }^{\dagger }\left( k\right) c_{p,\sigma ^{\prime
}}^{\dagger }\left( k^{\prime }\right) c_{p,\sigma ^{\prime }}\left(
k^{\prime }+q\right) c_{p,\sigma }\left( k-q\right)  \label{g-ology}
\end{eqnarray}
In this notation then, the space of parameters is closed under the
renormalization-group (RG) induced flow.

The Hubbard Hamiltonian and the above {\em g-}ology Hamiltonian clearly
differ since the dispersion relation is linearized and {\em g-}ology
contains fewer scattering terms than the original Hubbard model. If one
assumes that the high-energy processes that were dropped do not influence
the Physics, then $g_{1}=g_{2}=g_{4}=U$ for the Hubbard model. It is
important to notice, however, that this identification is an approximation.
In fact the {\em g-}ology Hamiltonian $H_{g}$ is a small-cut-off limit of
the Hubbard model and is strictly related to the Hubbard model only in a RG
sense. Generally, the appropriate initial values of the coupling constants
entering the RG would be such that, for example, $g_{4}\neq U$. Here by
rotational invariance parallel and perpendicular components of the coupling
constant $g_{1}$ are identical, a property that is preserved by the RG
transformation on the {\em g-}ology Hamiltonian.

\section{Quantum Monte Carlo results}

We are interested in the quarter-filled case, which corresponds to a large
class of organic conductors. Earlier results for the uniform magnetic
susceptibility $\chi $ include the following. First, zero temperature
results that have been obtained from the Bethe ansatz by Shiba\cite{Shiba}
and are shown for $U/t=2$, and $U/t=4$ as the left-most points on Fig 1.
Recently, J\"{u}ttner {\it et al.}\cite{Juttner} used a new approach based
on the Trotter-Suzuki mapping and a subsequent investigation of the quantum
transfer matrix to obtain the temperature-dependent results shown by the
solid line in Fig.1.%
\begin{figure}%
%
\centerline{\epsfxsize 12cm \epsffile{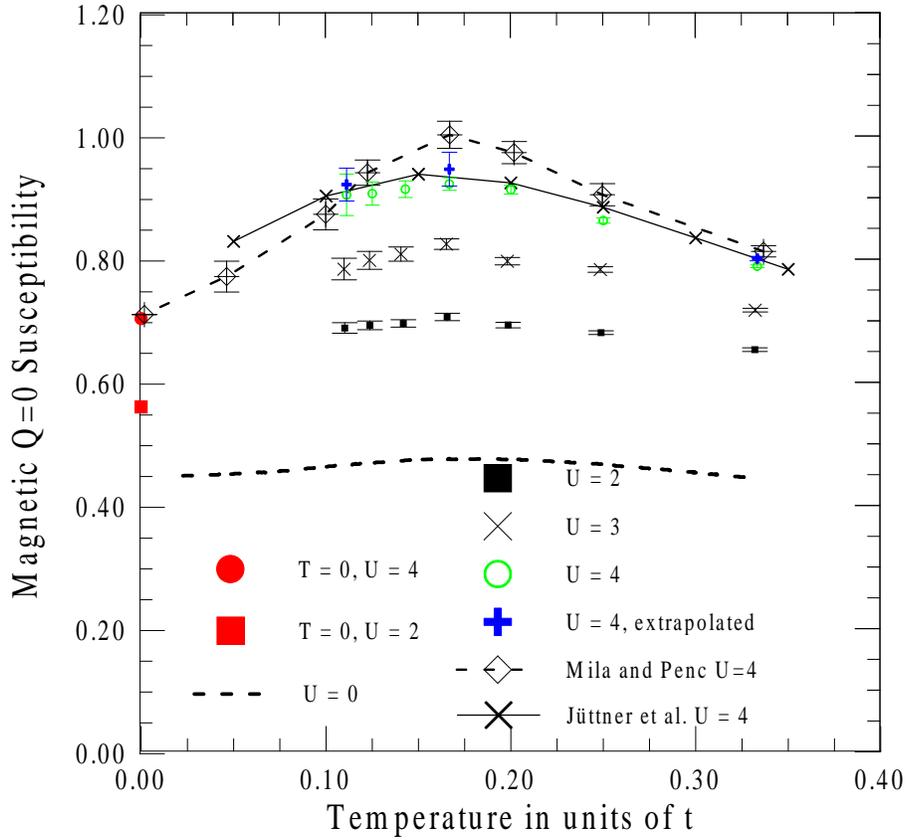}}%
%
\caption{Monte Carlo simulation results for the temperature dependent 
susceptibility as defined by Eq.(\ref{DefS}) and $T\chi =S\left( q=0\right) $. 
Our Monte Carlo results ($2\times 10^{5}$ measurements) are shown by 
symbols with error bars for different values of $U.$ For $U=4$ the extrapolation of 
our results to $\Delta \tau =0$ is also shown for three temperatures$.$
They are, within error bars, in agreement with the quantum transfer matrix results
of J\"uttner {\it et al. }\protect\cite{Juttner} shown by the
solid line. Points joined by the dashed line are from Ref.\protect\cite{Mila}
while the zero-temperature results shown by symbols are
from Shiba\protect\cite{Shiba}.}%
%
\label{fig1}%
%
\end{figure}%
%
\quad Earlier results had been obtained using a world line algorithm by Mila
and Penc.\cite{Mila} The corresponding results at $U/t=4$ are linked by a
dashed line in Fig.1. These results disagree substantially from those of
J\"{u}ttner {\it et al.\cite{Juttner} }As we show below, our Monte Carlo
simulations confirm the latter results. The calculations of Ref. \cite{Mila}
are probably less accurate because they were obtained not only from the
usual extrapolations to $\Delta \tau =0$ and $L=\infty $, they also required
extrapolation to $q=0$ of the finite wave vector results because the
world-line algorithm is for the canonical ensemble. Finite-temperature DMRG
calculations\cite{Moukouri} cannot yet reach sizes large enough to be
compared with the results of Fig.1, although for larger values of the
interaction it would be possible.

We have used the so-called determinantal Monte Carlo method (BSS algorithm) 
\cite{BSS} to obtain the other finite-temperature results in Fig.1 for $%
U/t=2,3,4$. Since equal-time quantities give better statistics, the quantity
that was computed is the magnetic structure factor for an $L$ site lattice 
\begin{equation}
S\left( q\right) =\frac{1}{L}\sum_{{\bf r}_{i},{\bf r}_{j}}\exp \left[
iq\left( r_{i}-r_{j}\right) \right] \left\langle \left( n_{i,\uparrow
}-n_{i,\downarrow }\right) \left( n_{j,\uparrow }-n_{j,\downarrow }\right)
\right\rangle .  \label{DefS}
\end{equation}
This quantity at $q=0$ is trivially related to the magnetic susceptibility
through the fluctuation-dissipation theorem $\left( S\left( 0\right) =T\chi
\right) $ . The results shown by isolated points with error bars in Fig.1
are for a $30$ site chain at quarter-filling. The units are chosen so that $%
t=1$. About $2\times 10^{5}$ measurements were taken for each point. For $%
U=4 $ and three temperatures, namely $T=1/3,$ $1/6$ and $1/9$, we have done
the extrapolation to $\Delta \tau =0$ using three values of $\Delta \tau .$
These results are plotted as crosses in Fig.1. Our extrapolated results are
in excellent agreement with those of J\"{u}ttner et al.\cite{Juttner}, while
our unextrapolated results systematically underestimate those of J\"{u}ttner 
{\it et al.}\cite{Juttner} by at most a few percent at the highest
temperatures. We conclude that the discretization step that we used in
imaginary time, $\Delta \tau =1/8,$ leads to a systematic underestimation
(of order \cite{Fye} $\left( \Delta \tau \right) ^{2}$) of the
susceptibility which is smaller than the statistical error on the figure
except perhaps for the two highest temperatures where the extrapolated
results are within two error bars of the unextrapolated ones.

As we shall see in the RG treatment, interactions cause a singularity in the
temperature derivative of the $q=0$ susceptibility at $T=0.$ Observation of
this singularity would require huge system sizes. However, at higher
temperature, where the susceptibility is regular, finite-size effects should
be negligible when the thermal de Broglie wavelength $\xi _{th}=v_{F}/\left(
\pi T\right) $ is smaller than the system size. For system sizes $30$ and
quarter-filling, this criterion means that finite-size effects should be
small for $T>0.02.$ In fact, for the lowest temperature we have considered $%
T\approx 0.1$, we have $\xi _{th}\approx 5$. We have verified for $T>0.1$
that indeed our results are the same, within statistical accuracy, for sizes 
$10,20,30,$ $40$.

The results shown in Fig.1 do not contain the factor $1/2$ for each external
spin vertex, hence they are larger by a factor of four than those defined in
the following section.

\section{Renormalization group approach}

The renormalization group provides a useful tool to understand the g-ology
Hamiltonian and its fixed point behavior, the Luttinger liquid. In the first
subsection, we summarize well known results for the renormalization group
flow of the parameters.\cite{Solyom79}\cite{Emery}\cite{Firsov}\cite
{BourbonnaisCaron} In the second subsection, we show how to apply
perturbative techniques to the small cut-off theory to compute the uniform
magnetic spin susceptibility.

\subsection{Renormalization group}

Following the Kadanoff-Wilson renormalization group procedure of Ref.\cite
{BourbonnaisCaron} we derive the one-dimensional scaling results that are
essential for the calculation of the magnetic susceptibility. One starts
with the partition function of the one-dimensional electron gas expressed in
terms of a functional integral over anticommuting fields, namely 
\begin{equation}
Z=\int \int D\psi ^{\ast }D\psi \!\;{\rm e}^{S[\psi ^{\ast },\psi ]}
\end{equation}
The Euclidean action $S=S^{0}+S_{I}$ corresponding to the {\em g-}ology
Hamiltonian (\ref{g-ology}), consists in a sum of free and interacting
parts. Using the definition

\begin{equation}
\psi _{p,\sigma }(k,ik_{n})=\sqrt{\frac{T}{L}}\sum_{j=1}^{L}\int_{0}^{\beta
}d\tau {\rm e}^{-ikr_{j}+ik_{n}\tau }\psi _{p,\sigma }(r_{j},\tau )
\end{equation}
with the fermionic Matsubara frequencies $k_{n}=\left( 2n+1\right) \pi T$, $%
n=0,\pm 1,\pm 2,...$ and the notations $\widetilde{k}=[k,k_{n}=(2n+1)\pi T]$
and $\widetilde{q}=(q,iq_{m}=2m\pi T)$ the two parts of the action are given
by 
\begin{equation}
S^{0}[\psi ^{\ast },\psi ]=\sum_{p,\widetilde{k},\sigma }G_{p}^{0-1}(%
\widetilde{k})\psi _{p,\sigma }^{\ast }(\widetilde{k})\psi _{p,\sigma }(%
\widetilde{k})
\end{equation}
and 
\begin{equation}
S_{I}[\psi ^{\ast },\psi ]=\frac{T}{2L}\sum_{\{\ p,\widetilde{k_{1}},%
\widetilde{k}_{2},\widetilde{q},\sigma \}}(g_{1}\delta _{\sigma _{1}\sigma
_{3}}\delta _{\sigma _{2}\sigma _{4}}-g_{2}\delta _{\sigma _{1}\sigma
_{4}}\delta _{\sigma _{2}\sigma _{3}})
\end{equation}
\[
\times \psi _{p,\sigma _{1}}^{\ast }\left( \widetilde{k_{1}}\right) \psi
_{-p,\sigma _{2}}^{\ast }(\widetilde{k_{2}})\psi _{-p,\sigma _{3}}(%
\widetilde{k_{2}}+\widetilde{q})\psi _{p,\sigma _{4}}(\widetilde{k_{1}}-%
\widetilde{q}) 
\]
\[
-\frac{T}{2L}\sum_{\{\ p,\widetilde{k_{1}},\widetilde{k}_{2},\widetilde{q}%
,\sigma ,\sigma ^{\prime }\}}g_{4}\ \psi _{p,\sigma }^{\ast }(\widetilde{k}%
_{1})\psi _{p,\sigma \prime }^{\ast }(\widetilde{k}_{2})\psi _{p,\sigma
\prime }(\widetilde{k}_{2}+\widetilde{q})\psi _{p,\sigma }(\widetilde{k}_{1}-%
\widetilde{q}), 
\]
To rewrite this equation, the $g_{1}$ term in Eq.(\ref{g-ology}) has been
subjected to the change of variable, $q\rightarrow k-k^{\prime }-2pk_{F}-q.$
In the usual {\em g-}ology terminology described above, the constants $g_{1}$
and $g_{2}$ stand for the backward and forward coupling constants between
right and left moving carriers whereas $g_{4}$ is the forward scattering
amplitude for electrons on the same branch. The sums over wave vectors in $%
S_{I}[\psi ^{\ast },\psi ]$ are restricted by the bandwidth cut-off $%
E_{0}=2v_{F}k_{0}$. The measure is given by 
\begin{equation}
D\psi ^{\ast }D\psi =\prod_{p,\sigma ,\widetilde{k}}d\psi _{p,\sigma }^{\ast
}(\widetilde{k})d\psi _{p,\sigma }(\widetilde{k})
\end{equation}
while the one-dimensional free propagator of the fermion field has the form 
\begin{equation}
G_{p}^{0}(\widetilde{k})=-\left\langle \psi _{p,\sigma }\left( \widetilde{k}%
\right) \psi _{p,\sigma }^{\ast }\left( \widetilde{k}\right) \right\rangle
_{0}=\frac{1}{ik_{n}-\epsilon _{p}\left( k\right) }
\end{equation}
with $\epsilon _{p}(k)$ as in Eq.(\ref{dispersion}).

In the bandwidth cut-off scheme, the renormalization group procedure in one
dimension consists in successive partial integrations of fermion degrees
freedom in the outer band-momentum shell corresponding to energies ${\frac 12%
}E_0(l)\geq \epsilon _p(k)>{\frac 12}E_0(l+dl)$ for electrons and $-{\frac 12%
}E_0(l)\leq \epsilon _p(k)<-{\frac 12}E_0(l+dl)$ for holes, with $%
E_0(l)=E_0e^{-l}$ an effective bandwidth cut-off at step $l\geq 0\quad
(E_0\equiv 2E_F)$. For each partial summation, a complete Matsubara
frequency sum is performed. Making use of the linked cluster theorem, the
partial trace can be formally written as

\begin{eqnarray}
Z &=&\int \int_{<}D\psi ^{\ast }D\psi {\rm e}^{S[\psi ^{\ast },\psi
]_{<}}\int \int D\overline{\psi }^{\ast }D\overline{\psi }e^{S_{0}\left[ 
\overline{\psi }^{\ast },\overline{\psi }\right] +S_{I}\left[ \overline{\psi 
}^{\ast },\overline{\psi },\psi ^{\ast },\psi \right] }  \nonumber \\
&=&Z_{\overline{0}}\int \int_{<}D\psi ^{\ast }D\psi {\rm e}^{S[\psi ^{\ast
},\psi ]_{<}}\exp \left( S[\psi ^{\ast },\psi ]_{<}+\sum_{n=1}^{\infty }%
\frac{1}{n!}\left\langle \left( S_{I}\left[ \overline{\psi }^{\ast },%
\overline{\psi },\psi ^{\ast },\psi \right] \right) ^{n}\right\rangle _{%
\overline{0},c}\right)   \label{Cluster}
\end{eqnarray}
Here the $\overline{\psi }^{\prime }s$ refer to fermion fields with a band
wave vector in the outer momentum shell while the $\psi ^{\prime }s$ pertain
to lower momentum degrees of freedom that are kept fixed. The outer shell
averages (connected diagrams) are defined by 
\begin{equation}
\langle \left( \overline{\psi }^{\ast }....\overline{\psi }\right) \rangle _{%
\overline{0},c}=Z_{\overline{0}}^{-1}\int \int D\overline{\psi }^{\ast }D%
\overline{\psi }\left( \overline{\psi }^{\ast }....\overline{\psi }\right)
e^{S^{0}[\overline{\psi }^{\ast },\overline{\psi }]}
\end{equation}
From Eq.(\ref{Cluster}), the two- and four-point vertices of the Euclidean
action are successively renormalized as a function of $l$. We can neglect
the effect of $g_{4}$ at this stage and reintroduce it later because it
leads to logarithmic terms only in higher order, as explained at the end of
this subsection. Taking into account the possible differences between
parallel and anti-parallel spin scatterings, three distinct four-point
vertices can renormalize. Upon renormalization one cannot tell the
difference between $g_{1\Vert }$ and $g_{2\Vert }.$ One is free to include
all the corresponding scattering processes in a quantity defined as $%
g_{1\Vert }\left( l+dl\right) -g_{2\Vert }\left( l+dl\right) $. Adding and
subtracting then a coupling $g_{2\Vert }\left( l+dl\right) =g_{2\bot }\left(
l+dl\right) $ to the renormalized action, one is left with a form that under
renormalization will preserve the structure of the original action, namely $%
g_{1\Vert }\left( l+dl\right) =$ $g_{1\bot }\left( l+dl\right) $ and $%
g_{2\Vert }\left( l+dl\right) =g_{2\bot }\left( l+dl\right) .$ In the
notation of Ref.\cite{BourbonnaisCaron}, one can write 
\begin{equation}
z_{1}^{-1}(l)G_{p}^{0}(\widetilde{k})\rightarrow
z_{1}^{-1}(l)z_{1}^{-1}(dl)G_{p}^{0}(\widetilde{k})
\end{equation}
\begin{equation}
z_{2,3}(l)\Gamma _{1,2}\rightarrow z_{2,3}(l)z_{2,3}(dl)\Gamma _{1,2}
\end{equation}
where the renormalization factors $z_{1,2,3}(dl)$ pick up the logarithmic
outer shell contributions (${\cal O}(dl)$) of the cumulant expansion in Eq.(%
\ref{Cluster}). If one performs the field rescaling $z_{1}^{\frac{1}{2}}\psi
^{(\ast )}\rightarrow \psi ^{(\ast )}$ in $S$ so that the free part of the
Hamiltonian is invariant, one gets the recursion relations for the coupling
constants 
\begin{equation}
g_{1,2}(l+dl)=z_{2,3}(dl)z_{1}^{-2}(dl)g_{1,2}(l)  \label{mult}
\end{equation}
At the one-loop level, there is no logarithmic one-particle self-energy
corrections so that $z_{1}=1$. The Hartree-Fock diagrams redefine the
chemical potential in such a way that Luttinger's Fermi surface\cite{Bedell}
is preserved. As for the four-point vertex renormalization factors, $%
z_{2}(dl)$ and $z_{3}(dl)$, they are obtained from the evaluation of the $n=2
$ outer-shell averages where 
\begin{equation}
S_{I}[\overline{\psi }^{\ast },\overline{\psi },\psi ^{\ast },\psi ]\sim \ 
\overline{\psi }_{p}^{\ast }\overline{\psi }_{-p}^{\ast }\psi _{-p}\psi
_{p}+\ \overline{\psi }_{p}^{\ast }\overline{\psi }_{-p}\psi _{-p}^{\ast
}\psi _{p}+...
\end{equation}
corresponding to outer-shell decompositions in the Cooper, and Peierls
channels. An essential characteristic of the one-dimensional electron gas
that emerges at the one-loop level is the quantum interference between both
channels. Thus when the logarithmic diagrams are evaluated at zero external
Peierls and Cooper variables, several diagrams cancel and only a $2k_{F}$
electron-hole bubble remains for the renormalization of $g_{1}$ whereas $%
g_{2}$ is affected by a Cooper ladder graph. The relevant diagrams appear in
Fig.2.\ 
\begin{figure}%
%
\centerline{\epsfxsize 12cm \epsffile{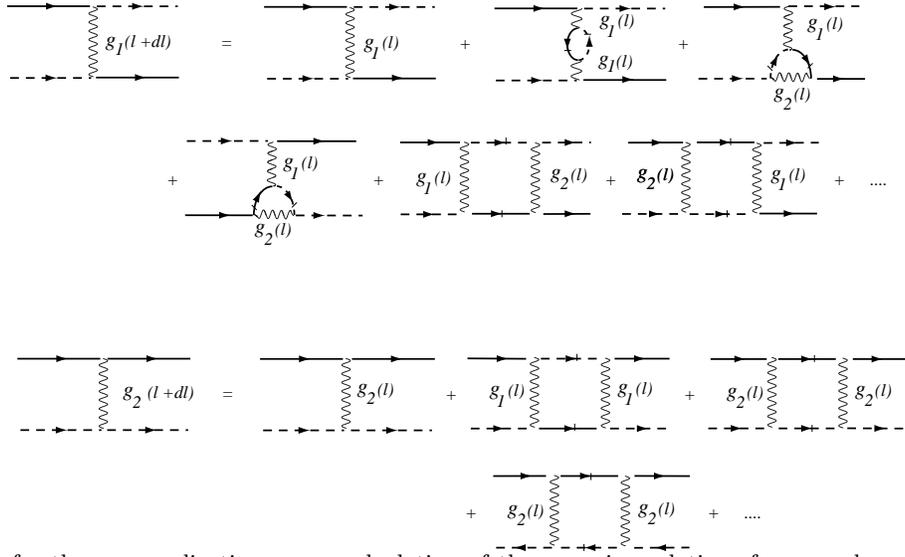}}%
%
\caption{Diagrams for the renormalization group calculation of the recursion relations 
for $g_{1}$ and $g_{2}$. Wiggly lines are for interactions, dashed (solid) lines 
are for fermions on the left $-k_{F}$ (right $k_{F}$) branch of the spectrum. 
The bar on a line indicates that the corresponding integration is in the 
small momentum shell being integrated.}%
%
\label{fig2}%
%
\end{figure}%
%
After the outer shell integration one is then left with, 
\begin{equation}
z_{2}(dl)=1-\widetilde{g}_{1}(l)dl
\end{equation}
\[
z_{3}(dl)=1-\frac{1}{2}\frac{\widetilde{g}_{1}^{2}}{\widetilde{g}_{2}}dl,
\]
where $\widetilde{g}_{1,2}\equiv g_{1,2}/\left( \pi v_{F}\right) $. From the
multiplicative renormalization (\ref{mult}), this gives the one-loop scaling
equations 
\begin{equation}
\frac{d\widetilde{g}_{1}}{dl}=-\widetilde{g}_{1}^{2}  \label{scaling}
\end{equation}
\[
\frac{d(2\widetilde{g}_{2}-\widetilde{g}_{1})}{dl}=0
\]
where the combination of coupling constants $2\widetilde{g}_{2}-\widetilde{g}%
_{1}$ is renormalization group invariant which is related to the
conservation of the particles on each branch for $g_{1}$ and $g_{2}$
scattering processes, when umklapp processes are neglected.

Another property of interest is that the decoupling of the above two scaling
equations can be understood as a consequence of the separation between spin
and charge long-wavelength degrees of freedom. This is clearly manifest when
the Hamiltonian representation $H_{I}(l)$ of the effective interacting part
of the action at step $l$ is written in the Landau channel in a rotationaly
invariant form 
\begin{equation}
H_{I}[g_{1}(l),g_{2}(l)]=\sum_{p,q}(2g_{2}-g_{1})\rho _{p}(q)\rho
_{-p}(-q)-g_{1}(l)\sum_{p,q}{\bf S}_{p}(q)\cdot {\bf S}_{-p}(-q).
\label{LandauH}
\end{equation}
Here the operators for the long-wavelength charge ($\rho _{p}$) and spin ($%
{\bf S}_{p}$) degrees degrees of freedom of branch $p$ are respectively
given, in operator form, by 
\[
\rho _{\pm p}(\pm q)=\frac{1}{2\sqrt{L}}{\sum_{k,\sigma }}^{\ast }{}c_{\pm
p,\sigma }^{\dagger }(k)c_{\pm p,\sigma }(k\pm q) 
\]
\begin{equation}
{\bf S}_{\pm p}(\pm q)=\frac{1}{2\sqrt{L}}{\sum_{k,\alpha \beta }}^{\ast
}{}c_{\pm p,\alpha }^{\dagger }(k){\vec{\sigma}}_{\alpha ,\beta }c_{\pm
p,\beta }(k\pm q).
\end{equation}
with the vector $\vec{\sigma}$ whose components are the usual Pauli
matrices. Because of the integrated degrees of freedom, the summations on
band wave vectors $k$ are restricted to the interval $\mid \epsilon
_{p}(k)\mid \leq E_{0}(l)/2.$ This way of writing the effective Hamiltonian
together with the scaling equations (\ref{scaling}) emphasizes that
interfering and logarithmically singular correlations of the Peierls and
Cooper channels do influence uniform correlations of the Landau channel.
This must be taken into account in the calculation of the uniform magnetic
susceptibility as we show shortly.

Before concluding this subsection, we mention how $g_{4}$ appears in the
partial trace operation Eq.(\ref{Cluster}). One can show that the coupling
of $g_{1}$ and $g_{2}$ to $g_{4}$ at higher order, also leads to logarithmic
corrections for $z_{2}$ and $z_{3}$, which are equivalent to a
renormalization of the Fermi velocity in the scaling equations (\ref{scaling}%
). One then has $\tilde{g}_{1}=g_{1}/(\pi v_{\sigma })$ and $2\tilde{g}_{2}-%
\tilde{g}_{1}=(2g_{2}-g_{1})/(\pi v_{\rho })$, where 
\begin{equation}
v_{\rho ,\sigma }=v_{F}[1\pm g_{4}(2\pi v_{F})^{-1}].  \label{Vitesses}
\end{equation}
Here $v_{\sigma }$ and $v_{\rho }$ are respectively the velocity of spin and
charge excitations. Thus for $g_{1}$ for example, one can write 
\begin{equation}
g_{1}(l)={\frac{g_{1}}{1+g_{1}(\pi v_{\sigma })^{-1}l},}  \label{g1(T)}
\end{equation}
while the charge coupling $2\tilde{g}_{2}-\tilde{g}_{1}$ remains an
invariant. As for the interaction term $g_{4}$ itself, it does not
renormalize at this order\cite{Solyom79}.

\subsection{Spin and charge susceptibilities from auxiliary fields.}

The renormalization described above is valid as long as the cut-off energy $%
E_{0}(l)/2$ is larger than the temperature. For smaller cut-off, the
contributions from loop integrations are not logarithmic because of the
Fermi occupation factors. While one could in principle modify the recursion
relations to account for this and continue their integration,\cite{Chitov}
it is simpler to apply a sharp cut-off procedure. We simply use the
effective action obtained for $l\approx \ln (E_{F}/T)$ and apply
perturbation theory to obtain the uniform susceptibility. We go through the
exercise of formally generating the perturbative result through the
auxiliary field (Hubbard-Stratonovich) method. This will allow us to exhibit
the accuracy of standard perturbative techniques for a linearized spectrum
in the Landau channel\cite{BourbonParis,Nelisse} and to find the infinite
slope of the susceptibility in the zero temperature limit. But first, we
must analyze the part of the Hamiltonian that we will take as the
unperturbed one.

\subsubsection{The $g_4$ theory}

Rescaling wave vectors to recover the original units, all band momenta $%
\left| k\right| $ are now smaller than $\overline{k}_{0}\approx T/v_{F}$
and, correspondingly, transfer momenta $\left| q\right| $ have to be of the
same order. Hence we are now working with a small cut-off theory involving
only low-frequency, low-momentum interactions within a single branch, 
\begin{equation}
H_{p}^{4}=\sum_{k,\sigma }\epsilon _{p}\left( k\right) c_{p,\sigma
}^{\dagger }\left( k\right) c_{p,\sigma }\left( k\right) +\frac{g_{4}}{2L}%
\sum_{k,k^{\prime }=-\overline{k}_{0}+pk_{F}}^{\overline{k}%
_{0}+pk_{F}}\sum_{q}\sum_{\sigma ,\sigma ^{\prime }=\uparrow }^{\downarrow
}c_{p,\sigma }^{\dagger }\left( k\right) c_{p,\sigma ^{\prime }}^{\dagger
}\left( k^{\prime }\right) c_{p,\sigma ^{\prime }}\left( k^{\prime
}+q\right) c_{p,\sigma }\left( k-q\right) .  \label{Hp}
\end{equation}
It will be useful then, later, to work in an interaction representation
where in the zeroth-order Hamiltonian $H_{p}^{4}$ the two branches $\pm
k_{F} $ do not interact with each other. We will need to know the value of
averages such as $\left\langle S_{p\alpha }\left( \widetilde{q}\right)
S_{p^{,}\alpha }\left( \widetilde{q}^{\prime }\right) \right\rangle _{4}$
computed with the above Hamiltonian. The label $\alpha $ refers to spatial
direction, $x,y,z,$ of the spin operator. Because the Hamiltonians $%
H_{p}^{4} $ are small cut-off Hamiltonians, the exact irreducible vertex in
the particle-hole channel can be taken as $g_{4}$ without further
Landau-theory-like renormalization from high-energy processes. A number of
cancellations occur for this model,\cite{Metzner93} so that the final
expression has an RPA like form, 
\begin{equation}
\left\langle S_{p\alpha }\left( \widetilde{q}\right) S_{p^{\prime }\alpha
}\left( \widetilde{q}^{\prime }\right) \right\rangle _{4}\equiv \delta
_{p,p^{\prime }}\delta _{\widetilde{q},-\widetilde{q}^{\prime }}\frac{1}{4}%
\frac{\chi _{p}^{0}\left( \widetilde{q}\right) }{1-\frac{1}{2}g_{4}\chi
_{p}^{0}\left( \widetilde{q}\right) }.  \label{Spin4}
\end{equation}
In this expression, the non-interacting susceptibility $\chi _{p}^{0}\left( 
\widetilde{q}\right) $ on one branch branch $p=\pm $ is given by 
\begin{equation}
\chi _{p}^{0}\left( \widetilde{q}\right) =-2\sum_{k}\frac{f\left[ \epsilon
_{p}\left( k\right) \right] -f\left[ \epsilon _{p}\left( k+q\right) \right] 
}{iq_{n}+\epsilon _{p}\left( k\right) -\epsilon _{p}\left( k+q\right) },
\end{equation}
where $f\left( \epsilon _{p}\left( k\right) \right) $ is the Fermi function.
At $iq_{n}=0,$ this is one half of the total bare susceptibility since the
sum over wave vectors is peaked near only one of the two Fermi points. At
small wave vector and zero temperature, 
\begin{equation}
\chi _{p}^{0}(q,iq_{n})=N(E_{F})\frac{pv_{F}q}{pv_{F}q-iq_{n}},
\label{SuscepBranche}
\end{equation}
with $N(E_{F})=(\pi v_{F})^{-1}$ the bare density of states per branch (half
of the total bare density of states). Similarly, the charge-charge
correlation function is given by\cite{Metzner93} 
\begin{equation}
\left\langle \rho _{p}\left( \widetilde{q}\right) \rho _{p^{\prime }}\left( 
\widetilde{q}^{\prime }\right) \right\rangle _{4}\equiv \delta _{p,p^{\prime
}}\delta _{\widetilde{q},-\widetilde{q}^{\prime }}\frac{1}{4}\frac{\chi
_{p}^{0}\left( \widetilde{q}\right) }{1+\frac{1}{2}g_{4}\chi _{p}^{0}\left( 
\widetilde{q}\right) }.  \label{Charge4}
\end{equation}
Note that in the $iq_{n}=0$, $q\rightarrow 0$ limit we have at low
temperature, 
\begin{equation}
\lim_{q\rightarrow 0}\left\langle S_{p\alpha }\left( q,0\right) S_{p\alpha
}\left( -q,0\right) \right\rangle _{4}=\frac{1}{4}\frac{(\pi v_{F})^{-1}}{%
1-g_{4}/\left( 2\pi v_{F}\right) }\equiv \frac{1}{4}\frac{1}{\pi v_{\sigma }}
\label{SqZero}
\end{equation}
\begin{equation}
\lim_{q\rightarrow 0}\left\langle \rho _{p}\left( q,0\right) \rho _{p}\left(
q,0\right) \right\rangle _{4}=\frac{1}{4}\frac{(\pi v_{F})^{-1}}{%
1+g_{4}/\left( 2\pi v_{F}\right) }\equiv \frac{1}{4}\frac{1}{\pi v_{\rho }},
\label{CqZero}
\end{equation}
where the spin and charge velocities are defined as above in Eq.(\ref
{Vitesses}).

\subsubsection{Auxiliary-field representation}

The magnetic susceptibility is obtained from a derivative with respect to an
external magnetic field ${\bf h}$. We choose units where the $g$ factor
times the Bohr magneton equals unity. In the presence of ${\bf h}${\bf ,}
the partition function takes the form, 
\begin{equation}
Z[{\bf h}]=Tr\left\{ e^{-\beta \left( H_{+}^{4}+H_{-}^{4}\right) }\,T_{\tau
}\,\exp \left[ -\int_{0}^{\beta }\left\{ H_{I}\left[ g_{1}(l),g_{2}(l),\tau %
\right] -\left( \sum_{p=\pm }{\bf S}_{p}(q,\tau )\right) \cdot {\bf h}%
(-q,\tau )\right\} \,d\tau \right] \right\} ,
\end{equation}
where we use the interaction representation in which the Hamiltonian $%
H_{p}^{4}$ studied in the previous section plays the role of the unperturbed
Hamiltonian. Using Gaussian integration (Hubbard-Stratonovich
decomposition), to decouple the interactions between branches, this may be
rewritten as 
\[
Z=Z_{4}\int \int {\cal D}\phi {\cal D}{\bf M}\,\exp \left(
-\sum_{q}\sum_{p}\int_{0}^{\beta }d\tau \,\left[ \phi _{p}(q,\tau )\phi
_{-p}(-q,\tau )+{\bf M}_{p}(q,\tau )\cdot {\bf M}_{-p}(-q,\tau )\right]
\right) 
\]
\begin{equation}
\times \left\langle T_{\tau }\exp \left( -\int_{0}^{\beta }\,d\tau {\cal H}%
\left[ \phi ,{\bf M},{\bf h},\tau \right] \right) \right\rangle _{4},
\label{HubbardStrat}
\end{equation}
where $\phi $ and ${\bf M}$ are real auxiliary fields for charge and spin
degrees of freedom respectively and 
\[
{\cal H}[\phi ,{\bf M},{\bf h},\tau ]=\sum_{p,q}\left[ 2i\sqrt{2g_{2}-g_{1}}%
\right. \,\rho _{p}(q,\tau )\phi _{p}(-q,\tau ) 
\]
\begin{equation}
+\left. 2\sqrt{g_{1}(l)}\,{\bf S}_{p}(q,\tau )\cdot {\bf M}_{p}(-q,\tau )-%
{\bf S}_{p}(q,\tau )\cdot {\bf h}(-q,\tau )\right]
\end{equation}
while $Z_{4}\equiv Tr\left[ e^{-\beta \left( H_{+}^{4}+H_{-}^{4}\right) }%
\right] $ and $\langle ...\rangle _{4}$ are respectively the partition
function and corresponding averages calculated with in the presence of the $%
g_{4}$ interaction only. As in previous sections, we use the following
definition for quantities in Matsubara frequencies $\omega _{m}=2\pi mT,$ $%
m=0,\pm 1,\pm 2,...$%
\begin{eqnarray}
{\bf h}\left( \tau \right) &=&\sqrt{T}\sum_{\omega _{m}}e^{-i\omega _{m}\tau
}{\bf h}\left( \omega _{m}\right) \\
{\bf h}\left( \omega _{m}\right) &=&\sqrt{T}\int_{0}^{\beta }d\tau
e^{i\omega _{m}\tau }{\bf h}\left( \tau \right) .
\end{eqnarray}

With these definitions, and the change of variable ${\bf M}_{p}\left( 
\widetilde{q}\right) \rightarrow {\bf M}_{p}\left( \widetilde{q}\right) +%
{\bf h}\left( \widetilde{q}\right) /\left( 2\sqrt{g_{1}}\right) $, $%
\widetilde{q}=(q,\omega _{m}=2\pi mT)$ the zero-field dynamic magnetic
susceptibility per unit length can be formally expressed in terms of an
average over the magnetic auxiliary field ${\bf M}$, that is 
\begin{eqnarray}
\chi _{\alpha }(\widetilde{q}) &=&\left\langle \sum_{p,p^{\prime
}}S_{p\alpha }\left( \widetilde{q}\right) S_{p^{,}\alpha }\left( -\widetilde{%
q}\right) \right\rangle =\left. \frac{\delta ^{2}\ln Z[{\bf h}]}{\delta
h_{\alpha }(-\widetilde{q})\delta h_{\alpha }(\widetilde{q})}\right| _{{\bf h%
}=0}  \label{Suscep} \\
&=&\frac{1}{g_{1}(l)}[\langle \sum_{p}M_{p,\alpha }(\widetilde{q}%
)\sum_{p^{\prime }}M_{p^{\prime },\alpha }(-\widetilde{q})\rangle -1] 
\nonumber
\end{eqnarray}
where the subscript $\alpha $ stands for the orientation of the magnetic
field.

Applying the linked-cluster theorem in Eq.(\ref{HubbardStrat}) and
evaluating the averages $\left\langle {\ldots}\right\rangle _{4}$ using
Eqs.(\ref{SqZero}) and (\ref{CqZero}) of the previous subsection allows one
to write, in the rotationally invariant case, 
\begin{equation}
Z=Z_{4}\int \int {\cal D}\phi {\cal D}{\bf M}\exp \left\{ -\sum_{\widetilde{q%
},p,p\prime }[\phi _{p}(\widetilde{q})A_{p,p\prime }(\widetilde{q})\phi
_{p\prime }(-\widetilde{q})+{\bf M}_{p}(\widetilde{q})B_{p,p\prime }(%
\widetilde{q}){\bf M}_{p\prime }(-\widetilde{q})]+{\cal O}\left( {\bf M}%
^{4},\phi ^{4},{\bf M}^{2}\phi ^{2}\right) \right\}  \label{ModeMode}
\end{equation}
with the matrix elements 
\begin{equation}
A_{p,p}(\widetilde{q})=\frac{1}{2}(2g_{2}-g_{1})\frac{\chi _{p}^{0}\left( 
\widetilde{q}\right) }{1+\frac{1}{2}g_{4}\chi _{p}^{0}\left( \widetilde{q}%
\right) }\equiv \frac{1}{2}(2g_{2}-g_{1})\chi _{c,p}^{0}\left( \widetilde{q}%
\right)
\end{equation}
\begin{equation}
A_{+,-}=A_{-,+}=1
\end{equation}
and 
\begin{equation}
B_{p,p}(\widetilde{q})=-\frac{1}{2}g_{1}\left( l\right) \frac{\chi
_{p}^{0}\left( \widetilde{q}\right) }{1-\frac{1}{2}g_{4}\chi _{p}^{0}\left( 
\widetilde{q}\right) }\equiv -\frac{1}{2}g_{1}\left( l\right) \chi _{\sigma
,p}^{0}\left( \widetilde{q}\right)  \label{B}
\end{equation}
\begin{equation}
B_{+,-}=B_{-,+}=1
\end{equation}
respectively for the charge and spin degrees of freedom of the Landau
channel.

In these expressions, $g_{1}\left( l\right) $ is given by Eq.(\ref{g1(T)})
with $l\approx \ln (\Lambda /T)$ where $\Lambda $ is a cut-off of the order
of the Fermi energy. hence, 
\begin{equation}
g_{1}\left( T\right) =\frac{g_{1}}{1+\frac{g_{1}}{\pi v_{\sigma }}\ln \frac{%
\Lambda }{T}}  \label{g*}
\end{equation}

The neglect of mode-mode coupling or anharmonic terms in Eq.(\ref{ModeMode})
is quite justified in one-dimension. Indeed we are in a low cut-off theory
so that only the $\widetilde{q}=0$ components of the Hubbard-Stratonovich
fields are important. When the unperturbed Hamiltonian is quadratic in
fermion fields, the coefficients of the $\widetilde{q}=0$ mode-mode coupling
terms vanish because they are derivatives of the density of states,\cite
{Hertz} a quantity that is a constant for a linear dispersion relation. For
the same reason, the spatial rigidity of correlations of the auxiliary
fields $\phi $ and ${\bf M}$ vanishes. Here the unperturbed part of the
Hamiltonian is the interacting $g_{4}$ theory. It is known that in the
theory with a linearized dispersion relation, the density fluctuations are
Gaussian, hence there is no mode-coupling term.\cite{Metzner93} Furthermore,
there is no singularity in this theory\cite{NoteSing}. Here we will take
into account the fact that the dispersion relation is not linear. This could
make mode-mode coupling terms become different from zero at high
temperature, far from the Luttinger liquid fixed point\cite{Dumoulin96}.

\subsubsection{Susceptibilities}

The Gaussian fluctuations of ${\bf M}$ evaluated with the functional Eq.(\ref
{ModeMode}) give us the magnetic susceptibility through Eq.(\ref{Suscep}).
We find in the rotationaly invariant case 
\begin{equation}
\chi (l)=\frac{1}{g_{1}(l)}\left( \frac{1}{2}\sum_{p,p^{\prime }}\left(
B^{-1}\right) _{p,p^{\prime }}-1\right)
\end{equation}
Using the expression Eq.(\ref{B}) for $B_{p,p},$ one obtains, 
\begin{equation}
\chi \left( \widetilde{q}\right) =-\frac{1}{4}\left[ \frac{g_{1}\left(
l\right) \chi _{\sigma ,+}^{0}\left( \widetilde{q}\right) \chi _{\sigma
,-}^{0}\left( \widetilde{q}\right) +\left( \chi _{\sigma ,+}^{0}\left( 
\widetilde{q}\right) +\chi _{\sigma ,-}^{0}\left( \widetilde{q}\right)
\right) }{\frac{1}{4}g_{1}^{2}\left( l\right) \chi _{\sigma ,+}^{0}\left( 
\widetilde{q}\right) \chi _{\sigma ,-}^{0}\left( \widetilde{q}\right) -1}%
\right]
\end{equation}
In the Luttinger-liquid limit, (linear dispersion relation) one finds for
the retarded spin susceptibility, 
\begin{eqnarray}
\chi ^{R} &=&-\frac{1}{4}\frac{2}{\pi v^{s}}\frac{\left( v^{s}q\right) ^{2}}{%
\left( \omega +i\eta \right) ^{2}-\left( v^{s}\widetilde{v}^{s}\right) q^{2}}
\\
v^{s} &=&v_{\sigma }+\frac{g_{1}(l)}{2\pi }=v_{F}-\frac{g_{4}}{2\pi }+\frac{%
g_{1}(l)}{2\pi } \\
\widetilde{v}^{s} &=&v_{\sigma }-\frac{g_{1}(l)}{2\pi }=v_{F}-\frac{g_{4}}{%
2\pi }-\frac{g_{1}(l)}{2\pi }
\end{eqnarray}
which reduces to the known result\cite{Solyom79},\cite{Metzner93} in the
limit $g_{1}=0.$

For the static susceptibility of interest to us, the result is simpler. In
that case, we take $\omega \rightarrow 0$ first and notice that 
\begin{equation}
\chi _{\sigma ,+}^{0}\left( q=0,\omega =0\right) =\chi _{\sigma
,-}^{0}\left( q=0,\omega =0\right) \equiv \chi _{p}^{0}\left( T\right) ,
\end{equation}
where $g_{1}\left( l\right) $ is evaluated at $l_{T}=\ln (E_{F}/T).$ This
leads to 
\begin{equation}
\chi (T)=\frac{\frac{1}{2}\left( \frac{\chi _{p}^{0}\left( T\right) }{1-%
\frac{1}{2}g_{4}\chi _{p}^{0}\left( T\right) }\right) }{1-\frac{1}{2}%
g_{1}(T)\left( \frac{\chi _{p}^{0}\left( T\right) }{1-\frac{1}{2}g_{4}\chi
_{p}^{0}\left( T\right) }\right) }  \label{SuscepChi}
\end{equation}
To leading order in $g_{1}(T)$, one finds 
\[
\chi (T)\approx {\frac{1}{2}}\chi _{\sigma }^{0}(T)[1+{\frac{1}{2}}%
g_{1}(T)\chi _{\sigma }^{0}(T)+...] 
\]
with 
\begin{equation}
\chi _{\sigma }^{0}(T)\equiv \frac{\chi _{p}^{0}\left( T\right) }{1-\left(
2\pi v_{F}\right) ^{-1}g_{4}}.
\end{equation}
In the absence of $g_{4}$, this expression, coincides with the result
obtained previously by Dzyaloshinskii and Larkin \cite{DL} and by Lee {\it %
et al.}\cite{Lee}. Our more general expression for the susceptibility Eq.(%
\ref{SuscepChi}) may be rewritten as 
\begin{equation}
\chi (T)={\frac{\frac{1}{2}\chi _{p}^{0}(T)}{1-{\frac{1}{2}}%
(g_{4}+g_{1}(T))\chi _{p}^{0}(T)}.}  \label{SuscepChi2}
\end{equation}
For the repulsive sector $g_{1}>0$, $g_{1}(T\rightarrow 0)\rightarrow 0$ is
irrelevant in the zero temperature limit and there one recovers the
Luttinger-liquid result\cite{Schulz}, 
\begin{equation}
\chi \left( T=0\right) =\left\langle \sum_{p,p^{\prime }}S_{p\alpha }\left(
0,0\right) S_{p^{,}\alpha }\left( 0,0\right) \right\rangle ={\frac{{1}}{2\pi
v_{F}[1-(2\pi v_{F})^{-1}g_{4}]}=}\frac{1}{2\pi v_{\sigma }}
\label{SuscepT=0}
\end{equation}

The analogous calculations for the charge fluctuations also lead to similar
results. It suffices to do the following substitutions in any of the above
spin susceptibility results: 
\begin{equation}
g_{4}\rightarrow -g_{4}
\end{equation}
\begin{equation}
g_{1}\rightarrow g_{1}-2g_{2}
\end{equation}
In particular, in the static limit, the charge susceptibility $\chi
_{c}\left( T\right) $ (or equivalently the isothermal compressibility $%
\kappa _{T}\left( T\right) =\chi _{c}\left( T\right) /n^{2}$) is given by 
\begin{equation}
\chi _{c}\left( T\right) =\left\langle \sum_{p,p^{\prime }}\rho _{p}\left(
0,0\right) \rho _{p}\left( 0,0\right) \right\rangle ={\frac{\frac{1}{2}\chi
_{p}^{0}}{1+{\frac{1}{2}}(g_{4}+2g_{2}-g_{1})\chi _{p}^{0}(T)}.}
\label{Compressibilite}
\end{equation}
Again in the repulsive sector $g_{1}>0$, $2g_{2}-g_{1}$ is a renormalization
group invariant when umklapp scattering can be neglected (cf. Eq. (\ref
{scaling})) and one recovers the Luttinger-liquid result in the
zero-temperature limit \cite{Schulz}, 
\begin{equation}
\chi _{c}\left( T=0\right) ={\frac{{1}}{2\pi v_{F}[1+(2\pi v_{F})^{-1}\left(
g_{4}+2g_{2}-g_{1}\right) ]}}
\end{equation}

\subsubsection{Infinite slope in the zero temperature limit and third law of
thermodynamics}

While in the absence of umklapp scattering the charge susceptibility comes
in the zero-temperature limit with zero slope, as in a Fermi liquid, the
dependence of the magnetic susceptibility Eq.(\ref{SuscepChi2}) on the
marginally irrelevant variable $g_{1}\left( T\right) $ implies an infinite
slope in the zero temperature limit. This can be seen as follows. Since $%
\partial \chi _{p}^{0}/\partial T$ has a finite limit as $T\rightarrow 0,$
the singular contributions comes from 
\[
\frac{\partial \chi (T)}{\partial T}\rightarrow \chi ^{2}(T)\frac{\partial
g_{1}(T)}{\partial T} 
\]
The temperature derivative of the marginally irrelevant variable equals
infinity at $T=0$ as can easily be obtained from the temperature derivative
of Eq.(\ref{g*}) 
\begin{equation}
\frac{\partial g_{1}(T)}{\partial T}=\frac{g_{1}}{\left( 1+\frac{g_{1}}{\pi
v_{\sigma }}\ln \frac{\Lambda }{T}\right) ^{2}}\frac{g_{1}}{\pi v_{\sigma }}%
\frac{1}{T}
\end{equation}
hence, 
\begin{equation}
\lim_{T\rightarrow o}\frac{\partial \chi (T)}{\partial T}=\infty .
\label{PenteChiT0}
\end{equation}

A superficial look at the zero temperature infinite slope Eq.(\ref
{PenteChiT0}) suggests a violation of the third law of thermodynamics.
Indeed, consider the grand potential $\Omega =E-TS-\mu N,$ whose
differential change is given by $d\Omega =-SdT-Nd\mu -{\bf M\cdot }d{\bf B.}$
Normalizing to unit volume on finds, 
\begin{eqnarray}
\left. \frac{\partial }{\partial T}\chi \right| _{T=0} &=&\left. \frac{%
\partial }{\partial T}\frac{\partial M}{\partial B}\right| _{T=0}=-\left. 
\frac{\partial }{\partial T}\frac{\partial ^{2}\Omega }{\partial B^{2}}%
\right| _{T=0} \\
&=&-\left. \frac{\partial ^{2}}{\partial B^{2}}\frac{\partial }{\partial T}%
\Omega \right| _{T=0}=\left. \frac{\partial ^{2}S}{\partial B^{2}}\right|
_{T=0}=\infty .
\end{eqnarray}

The infinite second derivatives of the entropy seem to contradict the third
law of thermodynamics that says that the entropy at zero temperature is
independent of external parameters. However, because the point $T=0,$ $B=0$
is a critical point (infinite correlation length), the free energy at this
point is not analytic and we are not allowed to invert the order of
differentiation as we did on the second line. Hence, one cannot conclude
that $\left. \frac{\partial }{\partial T}\chi \right| _{T=0}=\infty $
violates the third law.\cite{Eggert} The entropy of the Luttinger liquid
does vanish at zero temperature, independently of $B$ and $\mu .$\cite
{Schulz}

\section{Comparisons with Monte Carlo simulations}

In this section, we shall compare the RG results first with the
zero-temperature exact results and then with all the finite-temperature
results exhibited in Fig.1. It is not yet possible to do simulations at low
enough temperature to confirm or not the existence of the infinite slope
seen in the RG approach in the $T\rightarrow 0$ limit since this also
requires huge system sizes. It should however be possible to verify that
regular extrapolation of finite $T$ results to the $T=0$ limit is not
possible. At higher temperature and larger interaction strengths,
non-logarithmic terms become important. We will show that it is possible to
estimate these. In two dimensions, it has been possible to explain the
complete temperature-dependent magnetic properties of the Hubbard model by
using diagrammatic approaches\cite{ChenLiBourbonnais},\cite
{BulutScalapinoWhite} far from half-filling or the Two-Particle
Self-Consistent approach\cite{Vilk},\cite{Vilk2} at arbitrary fillings. We
briefly explore the predictions of these approaches in one-dimension. The
reason for their failure in one dimension will help understand the correct
way to proceed.

\subsection{Comparisons with the Renormalization Group approach}

To compare numerical Monte Carlo results with our RG results for the
susceptibility Eq.(\ref{SuscepChi}) or equivalently Eq.(\ref{SuscepChi2}),
(remember the trivial factor of four to compare with simulations) one needs
to know the initial values of $g_{1}$ and $g_{4}$ entering the scaling
equations. As argued above, it is not strictly correct for the Hubbard model
to assume that these constants can be taken as $g_{1}=g_{4}=U$ because the
dispersion relation is not linear, and the cut-off in the initial model is
not as in the {\em g-}ology model. Let us start by a comparison with the
zero temperature exact results of Shiba. We find that our result Eq.(\ref
{SuscepT=0}) differs from that of Shiba by at most $15\%$ up to $U=4$ if we
choose $g_{4}=U.$ At $U=2,$ the RG result is accurate to about $1\%.$ The
same conclusions were reached a long time ago in Ref.\cite{HirschScalapino}.
Hence, we conclude that at small coupling the estimate $g_{1}=g_{4}=U$
should be accurate.

To do a more general comparison at finite temperature, our results Eq.(\ref
{SuscepChi2}) and (\ref{g*}) require in addition a value for the cut-off $%
\Lambda .$ Taking $\Lambda =2,$ which corresponds roughly to half the
bandwidth, of the order of the Fermi energy, produces the results in Fig.3.%
\begin{figure}%
%
\centerline{\epsfxsize 6cm \epsffile{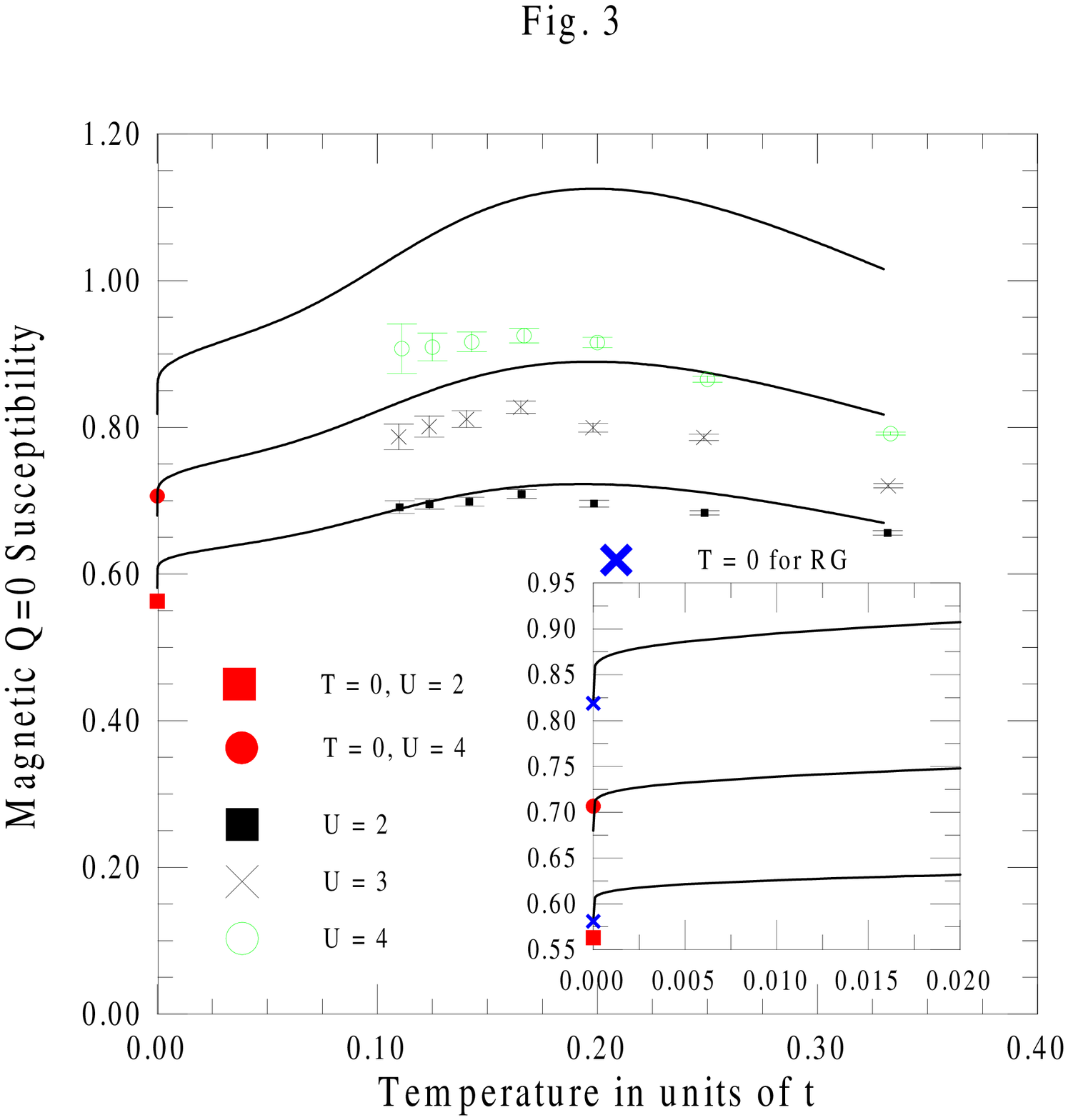}}%
%
\caption{Magnetic susceptibility $\chi $ evaluated from 
Eqs.(\ref{SuscepChi}) or equivalently Eq.(\ref{SuscepChi2}) 
with the naive replacements $g_{4}=$ $g_{1}=U$ and, using 
Eq.(\ref{g*}), 
$g_{1}\left( T\right) =g_{1}/\left( 1+\frac{g_{1}}{\pi v_{\sigma }}
\ln \frac{\Lambda }{T}\right) $ with $\Lambda =2$. 
The spin velocity is obtained from Eq.(\ref{Vitesses}), 
$v_{\rho ,\sigma }=v_{F}[1\pm g_{4}(2\pi v_{F})^{-1}]$. 
The inset shows the low temperature region. In the inset, 
the X symbols show the low-temperature limit predicted by the RG.}%
%
\label{fig3}%
%
\end{figure}%
%
\quad At low temperature, the results are not too sensitive to the value of $%
\Lambda $ which enters only logarithmically. The comparison in Fig.3 shows
that the disagreement with Monte Carlo results does become larger as $U$
increases. Higher order logarithmic terms will not change the picture since
they are smaller. As shown in the following subsections, the main source of
discrepancy resides in non-logarithmic contributions that are actually not
negligible at high temperature where the non-divergent (non-logarithmic)
contributions dominate the magnetic susceptibility.

Le us discuss in turn the various features appearing in Fig.3. The Monte
Carlo data shows that the temperature at which the maximum susceptibility
occurs is essentially independent of the interactions. The interactions
determine only the overall enhancement and sharpness of the maximum but not
the position. The position of the maximum $T_{\max }$ within RG also does
not depend strongly on interactions but there can be a $10\%$ shift in $%
T_{\max }$ when one changes the cut-off from $\Lambda =1$ to $\Lambda =2.$
However, since the RG results do not seem very reliable near the maximum, we
postpone this discussion to Sec.V.E.

Decreasing the temperature from the position of the maximum, the RG predicts
a first inflection point located at $T_{i}=0.099\pm 0.002.$ This temperature
is just at the limit of our Monte Carlo data. The inflection point can also
not be seen in the more recent data of Ref.\cite{Juttner} since only the
points $T=0.1$ and $T=0.05$ are available in the low-temperature region (See
Fig.1). This inflection point is again an effect that is caused by the band
structure and within the above accuracy the position is independent of
interactions. Indeed, while the curvature of the non-interacting
susceptibility $\partial ^{2}\chi _{0}/\partial T^{2}$ is negative at the
maximum, it must be positive at low temperature because of the
characteristics of the one-dimensional band with a parabolic bottom. The
position of this inflection point is the same, within the quoted accuracy,
whether one uses $\Lambda =1$ or $\Lambda =2$ or a temperature independent
interaction as in Sec.V.E. With a parabolic band instead of a cosine band,
one finds an inflection point at $T_{i}=0.101\pm 0.002$, again essentially
independent of cut-off or interactions.

Since the zero temperature value of the susceptibility is reached from
finite temperature with an infinite slope, as discussed in the previous
section, there is a second low-temperature inflection point. For $\Lambda
=2, $ it appears at $T_{L}=0.0329,$ $0.0328$, $0.0312$ for, respectively, $%
U=2,3,4.$ This temperature is roughly three times lower than our lowest
temperature in the Monte Carlo simulations and also lower than the other
lowest available numerical results exhibited in Fig.1. The infinite slope in
the susceptibility predicted by the RG appears confined to a small
temperature range even when seen on a magnified scale, as one can check in
the inset of Fig.3.

Contrary to the previous $T_{\max }$ and $T_{i},$ identified above, the
location of the low-temperature inflection point $T_{L}$ is not purely a
band structure effect. It is clearly absent from $\chi _{0}\left( T\right) .$
The low-temperature inflection point appears in the RG calculation because
of the competition between the band-structure effects that lead to $\partial
^{2}\chi _{0}/\partial T^{2}>0$ and the logarithmic singularity in $%
g_{1}\left( T\right) $ that leads to negative curvature in the full
susceptibility at low temperature. As long as the interaction is
sufficiently strong, this low-temperature inflection point occurs at a
temperature that is remarkably independent of interactions. For lower values
of the interaction, $U=0.5$ and $U=1,$ one finds that the inflection point
is at $T_{L}=0.0208$ and $T_{L}=0.0288.$ Comparing with the above results,
one sees that for $1<U<4$, the low-temperature inflection point is
independent of interaction within about $10\%$ ($T_{L}=0.031\pm 0.002)$
while for $0.5<U<4$ the position varies from $0.02$ to $0.03.$ The weak
dependence on interaction comes from several factors: a) The interactions
influence the susceptibility weakly, mainly to first order in the expansion
of the RPA-like denominators. b) The inflection point occurs in a regime
where $g_{1}\left( T\right) $ takes its asymptotic form, $\pi v_{\sigma
}/\ln \left( \Lambda /T\right) $ which depends on the interactions only
weakly through $v_{\sigma }.$ c) Only the order $T^{2}$ of the Sommerfeld
expansion of $\chi _{0}$ suffices to obtain an accurate result. These three
simplifications allow one to obtain an accurate analytical expression for
the location of the inflection point. However, since terms up to order $%
T^{2}\ln ^{3}\left( \Lambda /T\right) $ must be kept, the resulting equation
is transcendental and must be solved numerically.

The location of the low-temperature inflection point $T_{L}$ clearly depends
on the value of the cut-off $\Lambda $. However, since the dependence is
logarithmic, the above results are not so sensitive to the precise value of
the cut-off. For example, for $\Lambda =1$ one finds $T_{L}=0.0219$, $0.0304$%
, $0.0349$, $0.0345$, $0.0324$ for, respectively, $U=0.5,1,2,3,4.$ These
values of $T_{L}$ are larger than the corresponding values for $\Lambda =2$
by at most $6\%.$ For a parabolic band and $\Lambda =2$, one finds $%
T_{L}=0.035$, $0.035$, $0.034$ for $U=2,3,4$ values that are at most $8\%$
larger than the corresponding results for the cosine band with the same $%
\Lambda =2.$

\subsection{Diagrammatic Kanamori approach}

From now on, we concentrate on the susceptibility at temperatures above the
low-temperature inflection point, where the RG singularities do not
contribute appreciably to the susceptibility. It was suggested long ago by
Kanamori that the interaction appearing in RPA expressions should be
renormalized. Following this idea, it was shown \cite{ChenLiBourbonnais}
that in two dimensions the renormalized $U_{rn}$ can be accurately computed
as follows: 
\begin{equation}
U_{rn}=\left\langle \frac{U}{1+U\chi _{pp}\left( {\bf Q,}iq_{n}=0\right) }%
\right\rangle .
\end{equation}
In this expression, the quantity $\chi _{pp}\left( {\bf Q,}iq_{n}=0\right) $
is the Cooper bubble for a total incident momentum ${\bf Q}$. The average is
over the values of ${\bf Q}$. Several different types of averages can be
done, namely over the whole Brillouin zone,\cite{ChenLiBourbonnais} or over
wave vectors corresponding to values of ${\bf Q=k+k}^{\prime }$ such that
both ${\bf k}$ and ${\bf k}^{\prime }$ are within an energy equal to the
temperature $T$ of the Fermi surface.\cite{Groleau} The latter type of
approximation is closer in spirit to Fermi-liquid theory\cite{Dare} and
gives overall better results.

When the one-dimensional version of the procedure of Ref.\cite
{ChenLiBourbonnais} is applied to the present one-dimensional case, the
agreement with the Monte Carlo results is apparently much better at high
temperature than with the scaling approach. The temperature dependence is
essentially correct and there is simply an underestimation of the overall
magnitude of the susceptibility by $3\%,4\%,$ and $9\%$ for, respectively, $%
U=2$ $,3,$ and $4$. The value of $U_{rn}$ that we find may be approximated
by 
\begin{equation}
U_{rn}\approx \frac{U}{1+U\lambda }
\end{equation}
with $\lambda \approx \left\langle \chi _{pp}\left( {\bf Q,}iq_{n}=0\right)
\right\rangle \approx 0.29$.

In two dimensions, the agreement with Monte Carlo simulations is much better
than found here and it is valid for all wave vectors, as long as the filling
is such that there is no zero-temperature phase transition. In one
dimension, the bare susceptibility at $q=2k_{F}$ diverges logarithmically
with temperature at any filling because of nesting. This means that at
sufficiently low temperature, RPA with a temperature and wave vector
independent $U_{rn}$ predicts a finite temperature phase transition at any
filling. This is prohibited by the Mermin-Wagner theorem. Sure enough, for
the values of $U$ studied here, this transition occurs at a temperature much
lower than those investigated above with Monte Carlo, but nevertheless, this
is a question of principle that cannot be overlooked.

\subsection{Two-particle self-consistent approach (TPSC)}

The TPSC approach\cite{Vilk}\cite{Vilk2} avoids any finite-temperature phase
transition in both one and two-dimensions. Hence, we may check wether this
gives a better agreement with Monte Carlo simulations than the previous
approach. The one-dimensional version of the theory can be summarized as
follows. One approximates spin and charge susceptibilities $\chi _{sp}$, $%
\chi _{ch}$ by RPA-like forms but with two different effective interactions $%
U_{sp}$ and $U_{ch}$ which are then determined self-consistently using sum
rules. Although the susceptibilities have an RPA functional form, the
physical properties of the theory are very different from RPA\ because of
the self-consistency conditions on $U_{sp}$ and $U_{ch}$. The necessity to
have two different effective interactions for spin and for charge is
dictated by the Pauli exclusion principle $\langle n_{\sigma }^{2}\rangle
=\langle n_{\sigma }\rangle $ which implies that both $\chi _{sp}$ and $\chi
_{ch}$ are related to only one local pair correlation function $\langle
n_{\uparrow }n_{\downarrow }\rangle $. Indeed, using the
fluctuation-dissipation theorem in Matsubara formalism and the Pauli
principle one can write:

\begin{equation}
\frac{1}{\beta N}\sum_{\widetilde{q}}\chi _{ch}(\widetilde{q})=n+2\langle
n_{\uparrow }n_{\downarrow }\rangle -n^{2}=\frac{1}{\beta N}\sum_{\widetilde{%
q}}\frac{\chi _{0}(\widetilde{q})}{1+\frac{1}{2}U_{ch}\chi _{0}(\widetilde{q}%
)},  \label{sumCharge}
\end{equation}
\begin{equation}
\frac{1}{\beta N}\sum_{\widetilde{q}}\chi _{sp}(\widetilde{q})=n-2\langle
n_{\uparrow }n_{\downarrow }\rangle =\frac{1}{\beta N}\sum_{\widetilde{q}}%
\frac{\chi _{0}(\widetilde{q})}{1-\frac{1}{2}U_{sp}\chi _{0}(\widetilde{q})},
\label{sumSpin}
\end{equation}
where $\beta \equiv 1/T$, $n=\langle n_{\uparrow }\rangle +\langle
n_{\downarrow }\rangle $, $\widetilde{q}=(q,iq_{n})$ with $q$ the wave
vectors of an $N$ site lattice, $iq_{n}$ Matsubara frequencies and $\chi
_{0}(\widetilde{q})$ the susceptibility for non-interacting electrons. The
first equalities in each of the above equations is an exact sum-rule, while
the last equalities define the TPSC approximation for $\chi _{ch}(\widetilde{%
q})$ and for $\chi _{sp}(\widetilde{q})$. In this approach, the value of $%
\langle n_{\uparrow }n_{\downarrow }\rangle $ may be obtained
self-consistently\cite{Vilk} by adding to the above set of equations the
relation $U_{sp}=g_{\uparrow \downarrow }(0)\,U$ with $g_{\uparrow
\downarrow }(0)\equiv \langle n_{\uparrow }n_{\downarrow }\rangle /\langle
n_{\downarrow }\rangle \langle n_{\uparrow }\rangle .$ As shown in Ref.\cite
{Vilk}, the above procedure reproduces both the Kanamori-Brueckner screening
described in the previous section as well as the effect of Mermin-Wagner
thermal fluctuations, giving a phase transition only at zero-temperature in
two dimensions. In two dimensions, there is however a crossover temperature $%
T_{X}$ below which the magnetic correlation length $\xi $ can grow
exponentially. Quantitative agreement with Monte Carlo simulations is
obtained\cite{Vilk} for all fillings and temperatures in the weak to
intermediate coupling regime $U<8t$. The equation for charge Eq.(\ref
{sumCharge}) is not necessary to obtain the spin structure factor.

In one dimension, the absence of phase transition in this theory at finite
temperature can be proven as follows. Near the temperature at which the
phase transition would occur in RPA, $\delta U\equiv U_{mf,c}-U_{sp}\approx
0 $, ($U_{mf,c}\equiv 2/\chi _{0}\left( 2k_{F},0\right) $) the $q=2k_{F}$
susceptibility at zero Matsubara frequency is becoming very large so that
the self-consistency relation Eq.(\ref{sumSpin}) can be approximated by 
\begin{equation}
2T\int \frac{dq}{2\pi }\frac{2}{U_{sp}\xi _{0}^{2}(\xi ^{-2}+q^{2})}%
=n-2\langle n_{\uparrow }n_{\downarrow }\rangle -C,
\end{equation}
where 
\begin{equation}
\xi _{0}^{2}\equiv \frac{-1}{2\chi _{0}\left( q\right) }\left. \frac{%
\partial ^{2}\chi _{0}\left( q,0\right) }{\partial ^{2}q}\right| _{q=2k_{F}}
\end{equation}
\begin{equation}
\xi \equiv \xi _{0}(U_{sp}/\delta U)^{1/2},
\end{equation}
and where the integral is for $q$ around $2k_{F}$ or $-2k_{F}$ and $C$
contains contributions from non-zero Matsubara frequencies and from
corrections to the Lorentzian approximation used for the $iq_{n}=0$
contribution. Then, 
\begin{equation}
\frac{4T}{U_{sp}\xi _{0}^{2}}\int \frac{dq}{2\pi }\frac{1}{\xi ^{-2}+q^{2}}%
\approx \frac{4T}{U_{sp}\xi _{0}^{2}}\xi =n-2\langle n_{\uparrow
}n_{\downarrow }\rangle +C .
\end{equation}
Since the right-hand side is a finite number, $\xi $ behaves as $U_{sp}\xi
_{0}^{2}/\left( 4T\right) $, becoming infinite at zero temperature only.
There is however nothing to prevent a zero-temperature phase transition in
the theory, so that it cannot describe accurately the one-dimensional
systems at very low temperature.

Comparisons with the Monte Carlo simulations reveal discrepancies of order $%
25\%$ for the case $U=4$. This means that contrary to the two-dimensional
case, this approach does not even reproduce the Kanamori-Brueckner result
described in the previous section. If we had taken $\left\langle n_{\uparrow
}n_{\downarrow }\right\rangle $ from Monte Carlo data in the sum rule Eq.(%
\ref{sumSpin}) instead of computing it self-consistently from the {\it ansatz%
} $U_{sp}=U\langle n_{\uparrow }n_{\downarrow }\rangle /\langle
n_{\downarrow }\rangle \langle n_{\uparrow }\rangle $ the results would have
been much better. In other words, the calculation of $\langle n_{\uparrow
}n_{\downarrow }\rangle $ self-consistently is an assumption that fails in
one dimension even more drastically than the RPA functional form with an
effective $U_{sp}$. The whole approach fails completely at low temperature.
Indeed $U_{mf,c}\equiv 2/\chi _{0}\left( 2k_{F},0\right) $ vanishes as $%
T\rightarrow 0$ while $\delta U\equiv U_{mf,c}-U_{sp}$ has to remain
positive according to the above arguments. This in turn implies that $%
U_{sp}\ $tends to zero at zero temperature which means that the uniform
magnetic susceptibility would not be enhanced at zero-temperature with this
theory, contrary to both exact and renormalization group results. One could
have hoped to use this approach to evaluate non-logarithmic contributions at
high temperature and inject them in the RG expression, but in fact the
logarithmic temperature dependence of the susceptibilities starts at rather
high temperature.

\subsection{General reason for the failure of higher-dimensional approaches}

All of the above approaches fail for several reasons. They do not take into
account the destructive interference between Cooper and Peierls channel and
they try to describe the whole $q$ dependence of the susceptibility with a
single wave-vector-independent effective interaction $U_{sp}$ or $U_{rn}$.
This is very different from the scaling theory which clearly shows that the
effective interaction near $q=0$ is different from that near $q=2k_F$.

The fact that the magnetic susceptibility in one-dimension cannot be
described with a single wave vector independent effective interaction $%
U_{sp} $ or $U_{rn}$ is illustrated in Fig.4a.%
\begin{figure}%
%
\centerline{\epsfxsize 6cm \epsffile{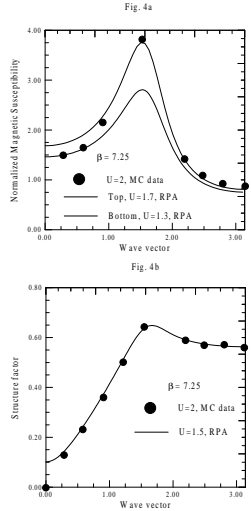}}%
%
\caption{a) Magnetic susceptibility $\chi \left( q\right) $ as 
a function of wave vector, normalized to its non-interacting 
$q=0$ value. The Monte Carlo data is from 
Ref.(\protect\cite{HirschScalapino}). The solid lines illustrate that when
the whole $q$ dependence is considered, RPA fits are inadequate, 
even with renormalized values of the interaction. 
b) Magnetic structure factor Eq.(\ref{DefS}) as a function of wave vector, 
$S\left( q\right) .$ The Monte Carlo data is from Ref.(\protect\cite{HirschScalapino}). 
The solid line illustrates that this time, as opposed to part (a), 
a simple RPA fit with a renormalized value of $U$ may work misleadingly well.}%
%
\label{fig4}%
%
\end{figure}%
%
\quad\ The simulations are taken from Ref.( \cite{HirschScalapino}). The
solid lines are for an RPA-like form 
\begin{equation}
\frac{\chi _{0}(q,0)}{1-\frac{1}{2}U\chi _{0}(q,0)}
\end{equation}
and the data is normalized by $\chi _{0}(q,0)$ as in Ref.(\cite
{HirschScalapino}). While the bare value for the simulation is $U=2$, one
needs a renormalized value $U=1.3$ to fit the components near $q=0$, while
to fit near $q=2k_{F}$ one needs $U=1.7$. Note also that the simulations are
done for the canonical ensemble so that the $q=0$ component is strictly zero
and is not shown.

The magnetic structure factor contains not only the above susceptibility,
but also all the non-zero Matsubara frequency components of the
susceptibility. That is why it is less sensitive to the $\pm 2k_{F}$ effects
of one dimension. As illustrated in Fig.4b, the Monte Carlo data of Ref.( 
\cite{HirschScalapino}) for $U=2$ can this time be fitted, misleadingly,
with a single renormalized $U=1.5$.

\subsection{Modified-Kanamori approach in one dimension}

In the renormalization group description of magnetic fluctuations, it is
clear that the effective interactions near $q=0$ and near $q=2k_{F}$ are
different. No finite or zero-temperature phase transition occurs even though
the uniform magnetic susceptibility is enhanced. However, this approach
takes into account only logarithmic terms and applies only in the vicinity
of either $q=0$ or $q=2k_{F}$. It does not allow one to draw the full $q$%
-dependent susceptibility appearing in Figs.4 for example. Furthermore, we
have seen that the renormalization group result for $q=0$ becomes inaccurate
compared with Monte Carlo simulations at high temperature. This can be
understood as follows. In addition to the fact that we have employed an
approximate sharp cut-off procedure, we do not know the exact value of $%
g_{4} $ and of $g_{1}$ entering the recursion relations. But more
importantly, logarithmic terms are less singular at high temperature so that
non-logarithmic terms also become important. To estimate non-logarithmic
contributions, we proceed as follows.

In the same spirit as the Kanamori approach in higher dimensions, we want to
find an effective interaction that contains the effect of other channels.
This time however, the effective interaction should be valid only for $q\sim
0$. The RG result suggests that it is the contribution from the
particle-hole channel that is important. Indeed, $g_{1}$ entering the
expressions (\ref{SuscepChi}) and (\ref{SuscepChi2}) for the susceptibility
contains the set of all diagrams generated by summing $2k_{F}$ electron-hole
bubbles.\cite{Nelisse} This can be seen from Fig.2 and from the recursion
relation Eq.(\ref{scaling}). In Fig.2, all cross terms involving $g_{1}g_{2}$
cancel each other in the cut-off theory so that only the $g_{1}^{2}$
contribution represented by the bubble is left. This shows that in the RG
approach only $2k_{F}$ electron-hole contributions are important. In the
computation of the magnetic susceptibility however, one can observe that
internal summations over all momentum transfers are present (including $q=0$
where electron-hole bubbles are also the only type of diagrams that
contribute in the calculation of $\chi $ for $H_{p}^{4}$ Refs.(\cite
{Solyom79,Metzner93})). In order to include these non-logarithmic effects in
the calculation of the susceptibility we average the resummed series of
electron-hole bubbles over the entire Brillouin zone. More specifically, we
take 
\begin{equation}
U_{m}=\left\langle {\frac{U}{1+U\chi _{eh}\left( q,iq_{n}=0\right) }}%
\right\rangle _{q},  \label{Um1}
\end{equation}
with 
\[
\chi _{eh}\left( q,iq_{n}=0\right) =-N^{-1}\sum_{k}{\frac{f\left[ \epsilon
\left( k-q\right) \right] -f\left[ \epsilon \left( k\right) \right] }{%
\epsilon \left( k-q\right) -\epsilon \left( k\right) }}. 
\]
Here the full effect of the lattice is restored by taking $\epsilon \left(
k\right) =2t\cos ka$ with the summation over $k$ that covers the entire
Brillouin zone $[-{\frac{\pi }{a}},{\frac{\pi }{a}}]$. It is clear from this
procedure that for $\left| q\right| \approx 0$, the average will contribute
to a $g_{4}$ type of process while for $\mid q\mid \approx 2k_{F}$ the
contribution will be to a $g_{1}$ type of process, as suggested by Eq.(\ref
{SuscepChi2}). One can also check that the Cooper and Peierls-type of bubble
diagram, such as those appearing in $g_{1}g_{2}$ processes in Fig.2, do
cancel at half-filling when averaged over the Brillouin zone. At this
filling, there is perfect particle-hole symmetry, even with a $2t\cos ka$
dispersion relation. Even if this cancellation is no longer strictly valid
away from half-filling, we write 
\begin{equation}
\chi (T)=\frac{\chi _{0}(q,0)}{1-\frac{1}{2}U_{m}\chi _{0}(q,0)}
\end{equation}
for the total susceptibility with $U_{m}$ computed from Eq.(\ref{Um1})

The results for the susceptibility are compared with the Monte Carlo results
in Fig.5.%
\begin{figure}%
%
\centerline{\epsfxsize 6cm \epsffile{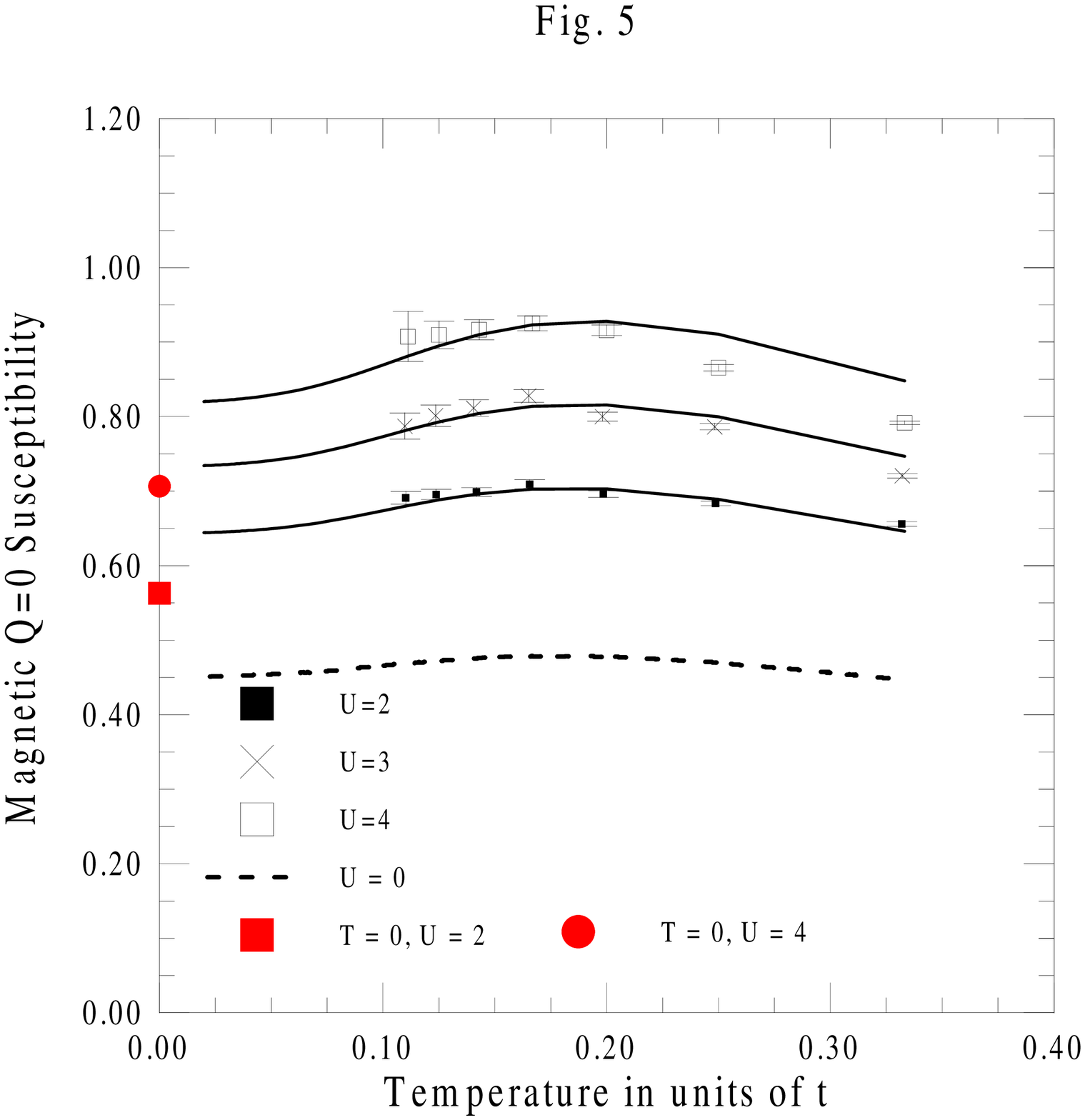}}%
%
\caption{Numerical results for the temperature dependent susceptibility, 
as already illustrated in Fig.1, compared with the Modified-Kanamori approach 
defined by an RPA from with a renormalized value of the interaction 
$U_{m}$ given as a function of the bare value $U$ by 
Eqs.(\ref{Um1}) or (\ref{Um2}).}%
%
\label{fig5}%
%
\end{figure}%
%
\ The agreement is within the statistical uncertainty except near $T=0.33$.
As in the Kanamori approach, for the range of temperatures illustrated in
Fig.5 we can approximate $U_{m}$ by a temperature-independent value given by 
\begin{equation}
U_{m}={\frac{U}{1+U{\lambda }_{m}}}  \label{Um2}
\end{equation}
with $\lambda _{m}=\left\langle \Lambda _{m}\left( q,iq_{n}=0\right)
\right\rangle _{q}.$ This last approximation gives, within about $2\%$ in
the worse case, the same value for the susceptibility as that obtained from
Eq.(\ref{Um1}). Using the low temperature value, $\lambda _{m}\simeq 0.25$,
over the full temperature range improves the agreement near $T=0.33$. At $%
T=0 $ the above modified-Kanamori approach overestimates the exact result by 
$13\%$ at $U=2$ and by as much as $17\%$ for $U=4$ showing clearly that a
Fermi-liquid like extrapolation of the finite-temperature data does not
yield the correct zero-temperature limit, as should be clear from Fig.5.
Independently of the RG then, it seems certain that a change in curvature is
needed to extrapolate to the correct zero-temperature limit. The RG result 
\cite{HirschScalapino} with $g_{4}=U$ on the other hand does have a change
in curvature below $T=0.1$ and extrapolates to the exact $T=0$ result\cite
{Shiba} to within $1\%$ for $U=2$ and makes physical sense even if $15\%$
deviations occur in the extrapolation around $U=4$.

Within the approach we just discussed, we can now come back to the question
of the location of the maximum in the spin susceptibility. The Monte Carlo
data suggests that the location of the maximum is essentially independent of
interaction, a result that is an obvious consequence of the analytical form
of the susceptibility that we used when $U_{rn}$ has a temperature
dependence that can be neglected. We find that $T_{\max }=0.178$ is
determined purely by the band structure$\ $and at quarter filling is even
the same at the $1\%$ level for a cosine or a parabolic band. Nevertheless,
we do not have a simple analytical expression for the position of the
maximum. Indeed, a Sommerfeld expansion of $\chi _{0}$ up to order $T^{6}$
predicts that $\chi _{0}$ always increases with temperature. It is the
competition of this increase with the eventual Curie-Weiss decrease of the
susceptibility at high temperature that produces the maximum. The
Curie-Weiss decrease is beyond the radius of convergence of the Sommerfeld
expansion.

To extract the strength of interactions from a measurement of $\chi \left(
T\right) $ one would need to know the bare value of $\chi _{0}\left(
T\right) .$ When $\chi _{0}\left( T\right) $ is unknown, one measure of the
strength of interactions that suggests itself, given the position of the
maximum $T_{\max }$ and of the high-temperature inflection point $T_{i}$ is $%
R=\left( \chi \left( T_{\max }\right) -\chi \left( T_{i}\right) \right)
/\chi \left( T_{\max }\right) .$ In percentage, one finds, for $U=2$ and $%
U=4,$ respectively, $R=4.2\%$ and $5.2\%.$ Unfortunately it turns out that
this result is too sensitive to the band structure to really be useful in
practice since for a parabolic band the above cosine band results are
replaced by $R=3.7\%$ and $4.5\%$.

\section{Discussion and conclusion}

We have shown in this paper that the temperature-dependent magnetic spin
susceptibility of the Hubbard model at quarter-filling has the following
general features in the weak to intermediate coupling regime $(0<U<4t).$ As
temperature is decreased, one encounters a maximum in the susceptibility at $%
T_{\max }\sim 0.18t$ that arises from the competition between the
Curie-Weiss high-temperature decrease and the low temperature increase
caused by the proximity to the characteristic van Hove singularity of
one-dimensional systems with a parabolic band bottom. Decreasing the
temperature, there is then an inflection point at $T_{i}\sim 0.1t.$ The
position of the maximum and of the inflection point are essentially
independent of interaction strength, as confirmed by Monte Carlo
simulations. Although a rough measure of the strength of the interactions
may be obtained by comparing the relative size of the susceptibility at the
maximum and at the inflection point this is not very reliable since one
needs high accuracy as well as rather detailed information on the band
structure: Indeed, a ratio $\left( \chi \left( T_{\max }\right) -\chi \left(
T_{i}\right) \right) /\chi \left( T_{\max }\right) $ of order $4\%$ may
correspond to either $U\sim 4t$ or $U\sim 2t$ depending on whether the band
structure is parabolic or cosinusoidal. As shown in Fig. 5, the magnetic
susceptibility curve obtained from Monte Carlo simulations between $T\sim
0.1t$ and $T\sim 0.33t$ may be reproduced very simply by an RPA-like form
with an effective interaction $U_{m}$ for the $q=0$ component of the
susceptibility$.$ For the case considered here, one finds $U_{m}={U/}\left(
1+U{\lambda }_{m}\right) $ with ${\lambda }_{m}\sim 0.25.$ That effective
interaction may be computed from the bare one using the subset of diagrams
that naturally appears in the RG calculation. This effective interaction
allows one to take into account non-logarithmic corrections that are beyond
the RG approach. The effective interaction is valid only near $q=0$.
Approaches to the interacting problem that work in dimensions larger than $%
d=1$ fail essentially because they keep an RPA form for all values of $q$,
an approximation that is incorrect in one-dimension because of the
destructive interference that occurs between $q=2k_{F}$ Peierls and $q=0$
Cooper channels.

At $T=0,$ the logarithmic terms in the leading marginally irrelevant
operator of the RG lead to an infinite slope $\partial \chi /\partial T.$
This Luttinger-liquid result is strikingly different from the Fermi liquid
prediction $\partial \chi /\partial T=0$ at $T=0$. Fig.5 clearly shows that
a Fermi-liquid extrapolation of the finite-temperature data to the $T=0$
exact result is inappropriate. The RG approach that we used also allows to
show that in the presence of band structure effects (parabolic band bottom)
the appearance of the Luttinger-liquid regime at low temperature is signaled
by a low-temperature inflection point located at $T_{L}=(0.031\pm .002)t$
for $t\leq U\leq 4t$ and $\Lambda =2t.$ Although the near independence of $%
T_{L}$ on the precise value of the interactions for sufficiently large bare $%
U$ does not have a simple origin, as discussed in Sec.V.A, it comes in large
part from the fact that in this case $T_{L}$ occurs in a regime where the
marginally irrelevant interaction takes its asymptotic form $\pi v_{\sigma
}/\ln \left( \Lambda /T\right) .$ This form depends on interactions only
weakly through $v_{\sigma }.$ The location of $T_{L}$ is also not so
sensitive to the (unkown) value of the cut-off since for $\Lambda =t$ one
finds that $T_{L}$ is larger than the corresponding values for $\Lambda =2t$
by at most $6\%$ over the whole range $0.5t\leq U\leq 4t.$ For $\Lambda =2t,$
$T_{L}$ can also be larger by at most $8\%$ for a parabolic band compared
with a cosine band. The above Luttinger-liquid regime $\left( \text{where }%
\partial ^{2}\chi /\partial T^{2}<0\right) $ occurs at a temperature lower
than what has been achieved by numerical calculations up to now. It is quite
a challenge to verify them. It is also important to realize that even though
the logarithmic Luttinger-liquid limit shows up in the spin susceptibility
at quite low temperature, logarithmic contributions do appear at higher
temperature in other quantities such as the longitudinal spin relaxation
time $T_{1}.$

Although our results are for the quarter-filled model, it is clear that all
the qualitative features should hold for fillings that are not too close to
half-filling. A good quantitative estimate for the location of the
characteristic features of the temperature-dependent spin susceptibility can
easily be obtained for any other filling using the simple analytical
expressions that we have found in Secs.IV.B and V.E. Our analytical
expressions may also be used to compute the charge susceptibility
(compressibility), which as we saw has no singularity in the $T\rightarrow 0$
limit, as long as umklapp scattering can be neglected\cite{umklapp}. One
could also use our approach to make quantitative predictions for more
general models than the Hubbard model. Comparisons with experiment should
help to establish appropriate bare microscopic parameters for the
Hamiltonians of one-dimensional organic conductors.

We are grateful to S. Moukouri for useful discussions and to G. J\"{u}ttner
for sending data. We acknowledge the support of the Natural Sciences and
Engineering Research Council of Canada (NSERC), the Fonds pour la formation
de chercheurs et l'aide \`{a} la recherche from the Government of Qu\'{e}bec
(FCAR), the Centre d'Applications du Calcul Parall\`{e}le de
l'Universit\'{e} de Sherbrooke for the use of an IBM-SP2.


\begin{references}
\bibitem{Revue}  C. Bourbonnais in {\it Strongly-interacting Fermions and
High-temperature Superconductivity, }Eds. B. Dou\c{c}ot and J. Zinn-Justin,
pp307-369 (North Holland, Amsterdam, 1995).{\it \ }

\bibitem{Wzietek93}  P. Wzietek and F. Creuzet and C. Bourbonnais and D.
J\'{e}rome and K. Bechgaard and P. Batail, J. Phys. I (France) {\bf 3},171
(1993).

\bibitem{Voit95}  J. Voit, Rep. Prog. Phys. {\bf 58,} 977 (1995).

\bibitem{DL}  I.E. Dzyaloshinskii and A.I. Larkin, Sov. Phys. JETP {\bf 34},
422 (1972).

\bibitem{Lee}  P.A. Lee, T.M. Rice, R. Klemm, Phys. Rev. B {\bf 15}, 2984
(1977).

\bibitem{BourbonParis}  C. Bourbonnais, J. Phys. I France {\bf 3}, 143
(1993).

\bibitem{BetheChi}  T. Usuki, N. Kawakami and A. Okiji, J. Phys. Soc. Japan 
{\bf 59}, 1357 (1990); N. Kawakami, T. Usuki and A. Okiji, Phys. Lett. A 
{\bf 137}, 287 (1989).

\bibitem{Juttner}  G. J\"{u}ttner, A. Kl\"{u}mper, J. Suzuki, Nucl. Phys. B 
{\bf 522}, 471 (1998).

\bibitem{Mila}  F. Mila and K. Penc, Phys. Rev B{\bf \ 51}, 1997 (1995).

\bibitem{Moukouri}  S. Moukouri and L.G. Caron {\it Low temperature
thermodynamics of the metallic Kondo lattice model }(unpublished).

\bibitem{Nelisse}  H. N\'{e}lisse, MSc Thesis, Universit\'{e} de Sherbrooke,
III-743, (1992).

\bibitem{Affleck94}  S. Eggert, I. Affleck and M. Takahashi, Phys. Rev.
Lett. {\bf 73}, 332 (1994).

\bibitem{Dumoulin96}  B. Dumoulin, C. Bourbonnais, S. Ravy, J.P. Pouget et
C. Coulon, 1996, Phys. Rev. Lett. {\bf 76}, 1360 (1996).

\bibitem{Kanamori}  J. Kanamori, Prog. Theor. Phys. {\bf 30,} 275 (1963).

\bibitem{Brueckner}  K.A. Brueckner and C.A. Levinson, Phys. Rev. {\bf 97},
1344 (1955); K.A. Brueckner and J.L. Gammel, Phys. Rev. {\bf 109,} 1023 and
1040 (1958).

\bibitem{ChenLiBourbonnais}  Liang Chen, C. Bourbonnais, T. Li, and A.-M.S.
Tremblay, Phys. Rev. Lett. {\bf 66}, 369 (1991).

\bibitem{Classiques}  I.E. Dzyaloshinskii and A.I. Larkin, Sov. Phys. JETP 
{\bf 34}, 422 (1972); Y. Bychkov, L.P. Gor'kov, and I.E. Dzyaloshinskii,
Sov. Phys. JETP {\bf 23}, 489 (1966).

\bibitem{Solyom79}  J. Solyom, Adv. Phys. {\bf 28}, 201 (1979).

\bibitem{Emery}  V.J. Emery, in {\it Highly Conducting one-dimensional Solids%
}, Eds. J.T. Devreese, R.P. Evrard, and V.E. van Doren (plenum, New York,
N.Y. 1979) p.247

\bibitem{Firsov}  Y.A. Firsov, V.N. Prigodin, Chr. Seidel, Phys. Rep. {\bf %
126}, 245 (1985).

\bibitem{BourbonnaisCaron}  C. Bourbonnais and L.G. Caron, Int. J. Mod.
Phys. B {\bf 5}, 1033 (1991); C. Bourbonnais and L.G. Caron, Europhys. Lett. 
{\bf 5,} 209 (1988); Physica B {\bf 143,} 451 (1986).

\bibitem{Shiba}  H. Shiba, Phys. Rev. B {\bf 6}, 930 (1972).

\bibitem{BSS}  R. Blankenbecler, D.J. Scalapino, et R.L. Sugar, Phys. Rev. 
{\bf D24}, 2278 (1981); J.E. Hirsch, Phys. Rev. B{\bf 31}, 4403 (1985); S.R.
White, D.J. Scalapino, R.L. Sugar, E.Y. Loh, J.E. Gubernatis, R.T.
Scalettar, Phys. Rev. {\bf B40}, 506 (1989); For a review, see E.Y. Loh, and
J.E. Gubernatis, in Electron Phase Transitions, p. 177-235, edited par W.
Hauke, Y.V. Kopaev (Elsevier, Amsterdam, 1992).

\bibitem{Fye}  M. Suzuki, Phys. Lett. {\bf 113A}, 299 (1985); R.M. Fye,
Phys. Rev. B {\bf 33,} 6271 (1986); R.M. Fye and R.T. Scalettar, Phys. Rev.
B {\bf 36}, 3833 (1987).

\bibitem{Bedell}  K. B. Blagoev, K. S.Bedell, Phys. Rev. Lett. {\bf 79},
1106 (1997).

\bibitem{Chitov}  G.Y. Chitov, and D. S\'{e}n\'{e}chal, Phys. Rev. B. {\bf 52%
}, 13487 (1995).

\bibitem{Metzner93}  W. Metzner and C. Di Castro, Phys. Rev. B {\bf 47} ,
16\ 107 (1993).

\bibitem{Hertz}  J.A. Hertz, Phys. Rev. B{\bf 14}, 1165 (1976).

\bibitem{NoteSing}  The singularity of the uniform susceptibility at $%
g_{4}/\left( 2\pi v_{F}\right) =1$ is beyond the weak coupling regime. Our
approach is not valid in the strong-coupling regime.

\bibitem{Schulz}  H.J. Schulz, Phys. Rev. Lett. {\bf 64}, 2831 (1990); Int.
J. Mod. Phys. B{\bf 5}, 57 (1991).

\bibitem{Eggert}  S. Eggert, oral communication.

\bibitem{HirschScalapino}  J.E. Hirsch and D.J. Scalapino, Phys. Rev. B {\bf %
27}, 7169 (1983).

\bibitem{BulutScalapinoWhite}  N. Bulut, D.J. Scalapino, and S.R. White,
Phys. Rev. B {\bf 47}, 2742 (1993).

\bibitem{Vilk}  Y.M. Vilk, Liang Chen, A.-M.S. Tremblay, Phys. Rev. B {\bf 49%
}, 13 267 (1994).

\bibitem{Vilk2}  Y.M. Vilk and A.-M.S. Tremblay, Europhys. Lett. {\bf 33},
159 (1996): Y.M. Vilk and A.-M.S. Tremblay, J. Phys. I France, {\bf 7}, 1309
(1997).

\bibitem{Groleau}  D. Groleau, MSc thesis III-729, Universit\'{e} de
Sherbrooke, 1992.

\bibitem{Dare}  See Appendix C of A.-M. Dar\'{e}, L. Chen, and A.-M.S.
Tremblay, Phys. Rev. B {\bf 49}, 4106 (1994).

\bibitem{umklapp}  When the band-filling is commensurate, {\it e.g. }at
quarter-filling, one expects logarithmic transients in the RG flow of the
charge coupling $2g_{2}-g_{1}.$ These transients should also lead to an
infinite slope at $T=0$ as for the magnetic susceptibility. These transients
get weaker in amplitude as the order of commensurability increases. See for
example T. Giamarchi, Physica B {\bf 230-232}, 975 (1997).
\end{references}
\end{document}